\documentclass[12pt]{article}

\usepackage{amsmath,amssymb,amsthm,latexsym} 
\usepackage{amsfonts}
\usepackage[english]{babel} 
\usepackage{graphicx,color} 

\usepackage{array}
\usepackage{mathtools}

\newtheorem{lemma}{Lemma}[section]

\newcommand{\bbone}{{\bf{1}}}

\newtheorem{theorem}{Theorem}[section]
\newcommand{\bea}{\begin{eqnarray}}
\newcommand{\eea}{\end{eqnarray}}
\newcommand{\bee}{\begin{equation}}
\newcommand{\ee}{\end{equation}}

\newcommand{\Tr}{{\rm Tr}}

\newcommand{\cC}{{\cal{C}}}

\newcommand{\cI}{{\cal{I}}}
\newcommand{\cR}{{\cal{R}}}

\newcommand{\cH}{{\cal{H}}}

\newcommand{\gF}{{\mathfrak{F}}}
\newcommand{\gT}{{\mathfrak{T}}}

\newcommand{\cS}{{\cal{S}}}
\newcommand{\LV}{\cS}
\newcommand{\cT}{{\cal{T}}}

\newcommand{\cF}{{\cal{F}}}

\newcommand{\prf}{{\noindent {\rm \bf Proof}\, }}

\def\spvecA#1;{\if;#1;\else #1\cr \expandafter \spvecA \fi}

\renewcommand{\theequation}{\Roman{section}-\arabic{equation}}
\renewcommand{\thesection}{\Roman{section}}

\begin{document} 

\title{Constructive Matrix Theory\\ for Higher Order Interaction}
\author{
Thomas Krajewski$^{\star}$, Vincent Rivasseau$^{\dagger}$, Vasily Sazonov$^{\dagger}$\\
\\
$^{\star}\; $Centre de Physique Th\'eorique, CNRS UMR 7332\\
Universit\'e Aix-Marseille,
F-13009 Marseille  \\
$^{\dagger}$ Laboratoire de Physique Th\'eorique, CNRS UMR 8627,\\ 
Universit\'e Paris-Sud,  F-91405 Orsay}

\maketitle 

\begin{abstract} 
This paper provides an extension of the constructive \emph{loop vertex expansion}
to stable matrix models with interactions of arbitrarily high order. We introduce a new representation
for such models, then perform a forest  expansion on this representation. It allows us 
to prove that the perturbation series of the free energy for such models is analytic
in a domain \emph{uniform in the size $N$ of the matrix}. 
\end{abstract} 

\begin{flushright}
LPT-20XX-xx
\end{flushright}
\medskip


\vfill\eject

\section{Introduction}

The loop vertex expansion (LVE) 
was introduced in  \cite{Rivasseau:2007fr} to provide a constructive method for quartic matrix models
\emph{uniform in the size of the matrix}. In its initial version
it combines an intermediate field representation
with replica fields and a forest formula \cite{BK,AR1} 
to express the free energy of the theory 
in terms of a convergent sum over trees. 
This loop vertex expansion in contrast with traditional constructive methods 
is \emph{not} based  on cluster expansions nor involves small/large field conditions.

\begin{itemize}

\medskip\item 
Like Feynman's perturbative expansion, the LVE allows to compute connected quantities at a glance: 
the partition function of the theory is expressed by a sum over forests, and its logarithm is exactly the same sum 
but restricted to \emph{connected} forests, i.e. \emph{trees}. This is simply  because the amplitudes factorize over
the connected components of the forest,

\medskip\item 
the functional integrands associated to each forest or tree are absolutely and \emph{uniformly} convergent 
for any value of the fields,

\medskip\item 
the convergence of the LVE implies Borel summability of the usual perturbation series and the LVE 
directly computes the Borel sum,

\medskip\item  the LVE is in fact conceptually an \emph{explicit repacking} of infinitely many 
subsets of pieces of Feynman amplitudes so that the packets provide a convergent rather than divergent expansion \cite{Rivasseau:2013ova}. 

\medskip\item 
in the case of combinatorial field theories of the matrix and tensor type \cite{Gurau:2011xp,Guraubook}, suitably rescaled 
to have a non-trivial $N \to \infty$ limit \cite{Hooft,Gurau:2010ba,Gurau:2011aq,Gurau:2011xq}, the
Borel summability obtained in this way is \emph{uniform} in the size $N$ of the model 
\cite{Rivasseau:2007fr,Gurau:2014lua,Gurau:2013pca,Delepouve:2014bma}.

\end{itemize}

The LVE method can be developed for ordinary field theories with cutoffs  \cite{MR1}. A multiscale version (MLVE) \cite{Gurau:2013oqa} 
can include renormalization 
\cite{Delepouve:2014hfa,Lahoche:2015zya,Rivasseau:2017xbk,Rivasseau:2016rgt,Rivasseau:2014bya}
\footnote{However models built so far are only of the superrenormalizable type.}.
This MLVE is especially adapted to resum the renormalized series
of non-local field theories of the matrix or tensorial type. For ordinary local field theories, and in contrast with the more traditional
constructive methods such as cluster and Mayer expansions, it is still until now less efficient in
providing the spatial decay of truncated functions. See however \cite{MR1} and \cite{Fang}.

There was until recently a big limitation of the method: it did apply only to quartic interactions.
Progress to generalize the LVE to interactions of higher order has been slow. 
It is possible to generalize the intermediate field representation to interactions of order higher than 4
\cite{Rivasseau:2010ke,LR1,Lionni:2016ush}, using \emph{several} intermediate fields. However these representations all
imply oscillating Gaussian integrals and lead for matrix or tensor models
to analyticity domains for the free energy which are not uniform 
in the size $N$  of the matrix or tensor \cite{LR1,Lionni:2016ush}.

In \cite{Rivasseau:2017hpg} a new representation, called
\emph{loop vertex representation} (hereafter LVR), was introduced in the simple case of scalar 
monomial interactions of arbitrarily high even order. It does not suffer from the previous defects and it uses the initial 
fields of the model rather than intermediate fields. It was 
found through selective Gaussian integration of one particular field per vertex, hence giving rise to a new kind of ``single loop" vertex similar to those found in the Gallavotti quantum field theoretic version of 
classical mechanics \cite{Gal} or in the quantum field theory formulation of the Jacobian conjecture \cite{Abdesselam:2002cy,deGoursac:2014ufa}, see also \cite{Abdesselam:2002} for an algebraic version of this formalism.
It was quickly noticed that this LVR representation is in fact a \emph{reparametrization} of the functional integrand
into a Gaussian one. The  single loop vertices form the natural expansion of the Jacobian of the transformation, which is a determinant. This is the deep reason for which the LVE applied to this new representation 
then works. Indeed a determinant has slow
``logarithmic" growth at large field. In particular  its partial derivatives are typically bounded. 
The LVE could never converge for the initial Bosonic interactions 
because it has unbounded derivatives at large field.

\begin{figure}[!ht]
\begin{center}
{\includegraphics[width=7cm]{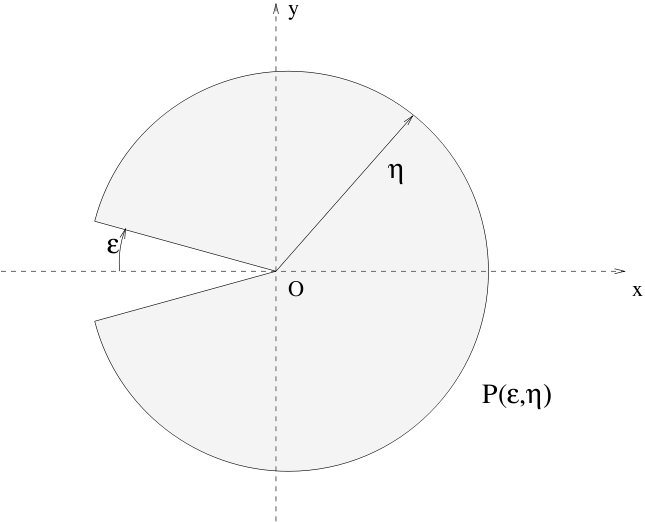}}
\end{center}
\caption{A pacman domain}
\label{pacman0}
\end{figure}

In this paper we apply the idea of reparametrization invariance to matrix models 
and essentially extend the results of \cite{Gurau:2014lua} to monomial interactions of arbitrarily high even order.
Our main result, the Theorem \ref{main} states that the free energy of such models is analytic
for $\lambda$ in an open  ``pacman domain" (see Figure \ref{pacman0})
\bee
P(\epsilon, \eta) := \{0<|\lambda |<\eta, \vert \arg \lambda \vert < \pi - \epsilon\}, \label{pacman}
\ee
with $\epsilon$ and $\eta$ positive and small numbers \emph{independent of the size $N$ of the matrix}. Extension of this theorem
to cumulants and a constructive version of the $1/N$ expansion are also consequences of the method left to the reader.
We intend also to explore links with the topological recursion approach to random matrices \cite{Eynard:2015aea}.

An unexpected difficulty of this paper compared to \cite{Rivasseau:2007fr}, \cite{Rivasseau:2017hpg} or 
\cite{Gurau:2014lua} is to deal with the non-factorization of the two sides of the ribbon loop
in a vertex of the loop vertex representation. Fortunately this difficulty can be
solved by using Cauchy holomorphic matrix calculus, which allows to factorize
the matrix dependence on the two sides of the ribbon, see Lemma \ref{lemmafactor}
below.
The price to pay is that one has to prove convergence of these contour integrals and this requires
a bit of convex analysis.

\medskip

The plan of our paper goes as follows.
In section II we introduce the LVR representation and its factorization through holomorphic calculus. 
In section III we perform the LVE on this representation. In Section IV
we establish the functional integral and contour bounds, completing the proof of Theorem \ref{main}. 
Four appendices gather some additional aspects: the first one is devoted to an alternative 
derivation of the LVR, the second one to an integral representation
of the Fuss-Catalan function that we need for the third one, devoted to the justification 
of the LVR beyond perturbation theory and the last appendix is devoted to its relationship to the ordinary perturbation theory.

\section{Effective action} 
Consider a complex matrix model with stable interaction of order $2p$, where $p\ge 2$ is an integer 
which is fixed through all this paper. The model has partition function
\begin{eqnarray}
Z(\lambda, N) &:=& \int\, dM dM^\dagger   \, e^{- N  S(M, M^\dagger )}, \label{Zsq}
\\
S(M, M^\dagger ) &:=&  \Tr \{ M M^\dagger +\lambda (M M^\dagger)^p\} .
\label{ZAsq}
\end{eqnarray}
$M$ is a random complex square matrix of size $N$ and 
the stable case corresponds to a positive coupling constant $\lambda$.
The goal is to compute the ``free energy"
\bee
F(\lambda, N) := \frac{1}{N^2} \log Z[\lambda, N] \label{Fsq}
\ee
for $\lambda$ in a domain independent of $N$.
The case of a rectangular $N_l \times N_r$ matrix is also important, as it allows
to interpolate between vectors and matrices,
and to better distinguish \emph{rows and columns}. We can introduce
the Hilbert spaces $\cH_l$ with  $\dim \cH_l = N_l$ and $\cH_r$ with $\dim \cH_r = N_r$. Remark that 
the two matrices  $M M^\dagger$ and  $M^\dagger M$ are distinct, the first one being $N_l$ by $N_l$ and the 
second $N_r$ by $N_r$, but crucially for what follows \emph{they have the same trace}, so all our computations
will be done involving only one of them, say $M M^\dagger$. In tensor products we may distinguish left and right factors;
for instance $A \otimes_{lr} B$ means an element of $\cH_{lr} := \cH_l \otimes \cH_r$, $\bbone_{lr}$ the identity in 
$\cH_{lr}$ and so on.
 For simplicity and without loss of generality
we can assume $N_l \le N_r$. Then, the partition function in the rectangular case is
\begin{eqnarray}
Z(\lambda, N_l,  N_r) &:=& \int\, dM dM^\dagger   \, e^{- N_r  S(M, M^\dagger )},\label{Zrect}
\\
S(M, M^\dagger ) &:=&  \Tr_l \{ M M^\dagger +\lambda (M M^\dagger)^p\} .
\label{ZArect}
\end{eqnarray}
and the quantity of interest is 
\bee
F(\lambda, N_l ,N_r) := \frac{1}{N_l N_r} \log Z(\lambda, N_l,N_r) . \label{Frect}
\ee
Of course there are similar formulas
using right traces $\Tr_r$. Also sources can be introduced to compute cumulants etc...

The standard perturbative approach to models of type \eqref{Zsq} or \eqref{Zrect} expands the exponential of the interaction 
into a Taylor series. However, polynomial interactions lead to divergent
perturbative expansions. To avoid this problem, we folllow the strategy of  \cite{Rivasseau:2017hpg}  
and first rewrite $Z[\lambda, N_l,  N_r]$ in another integral representation, 
called the \emph{loop vertex representation} (LVR), in which the interaction
\emph{grows only logarithmically} at large fields. One of the key elements of the
LVR construction is the Fuss-Catalan function $T_p$ \cite{FussCatalan}
defined to be the solution analytic at the origin  of the algebraic equation
\begin{equation}
z T_p^p(z) - T_p(z) + 1 = 0\, .
\label{FCEq}
\end{equation}
For any square matrix $X$ we also define the matrix-valued function
\bea
A(\lambda,X)&:=& X T_p(- \lambda X^{p-1})\, \label{Adef}
\eea
so that from \eqref{FCEq}
\begin{equation}
X = A(\lambda,X) + \lambda A^p (\lambda ,X)\,.
\label{AEq}
\end{equation}
We often write simply $A(X)$ for $A(\lambda,X)$ when no confusion is possible.
Finally we define an $N_l$ by $N_l$ square matrix $X_l$ and an $N_r$ by $N_r$ square matrix $X_r$ through
\bea X_l &:=&  M M^\dagger, \quad X_r :=  M^\dagger M . \label{defX} 
\eea

The loop vertex representation is
then given by
\begin{theorem} \label{theoremlvr} In the sense of \emph{formal power series in} $\lambda$
\begin{eqnarray}\nonumber
Z ( \lambda, N_l,  N_r ) &=& \int\, dM dM^\dagger\, \exp\{- N_r \Tr_l X_l + \LV(X_l, X_r) \}
\label{Zgood}
\end{eqnarray}
where $\LV$, the \emph{loop vertex action} is  
\begin{eqnarray}\nonumber
\LV(X_l, X_r)
&=& -  \Tr_{lr}\log\Big[\bbone_{lr} + \lambda \sum_{k = 0}^{p-1} A^k(X_l) \otimes_{lr} A^{p-1-k}(X_r)
  \Big].\\
\label{ZAgood}
\end{eqnarray}
In \eqref{ZAgood} the $N_l$ by $N_l$ matrix $A^k(X_l)$ acts on the left index of $\cH_{lr}$
and the $N_r$ by $N_r$ matrix $A^{p-1-k}(X_r)$ acts on the right index of $\cH_{lr}$.
\end{theorem}
\proof
Remark first that this formula exactly coincides with equations (II.12) and (II.16) of  \cite{Rivasseau:2017hpg} 
in the scalar case  $N_l = N_r =1$. We work first at the level of formal power series in order
not to worry about convergence. However Theorem \ref{theoremlvr}  \emph{holds beyond formal power series}
as proved in Appendix \ref{CorrVCh}.

Since \eqref{ZAgood} is crucial for the rest of the paper 
we propose two different proofs. The first one, below, relies on a change of variables on $M$
and the computation of a Jacobian\footnote{This change of variables is in fact well-defined on the eigenvalues of 
$X_l$ and $X_r$, and the unitary group part plays no role.}. A second perhaps more concrete proof
relies as in  \cite{Rivasseau:2017hpg} on Gaussian integration and will be given in Appendix \ref{partinteg}. 

We perform a change of variables $M \to P$ where $P$ is again an $N_l$ by $N_r$ rectangular matrix.
We write 
\bea Y_l &:=& P P^\dagger, \quad Y_r := P^\dagger P 
\label{YPch}
\eea
and define $P(M) $ (up to unitary conjugation) through the implicit function formal power series equation
\bea
X_l &:=& A(Y_l), \quad X_r := A(Y_r).
\label{XAYch}
\eea
Thanks to \eqref{AEq}, the action transforms to
\begin{eqnarray}
S(M, M^\dagger ) = 
\Tr_l  (X_l + \lambda X_l^p) =   \Tr_l [A(Y_l) +\lambda A^p (Y_l)] =  \Tr_l Y_l .
\end{eqnarray}
hence it becomes the ordinary Gaussian measure on $P, P^\dagger$.
The new interaction lies therefore entirely in the \emph{Jacobian} of the $M \to P$ transformation.
The transformation \eqref{YPch}, \eqref{XAYch} can be written more explicitly as
\begin{eqnarray}
\nonumber
  M &:=& P P^\dagger T_p(-\lambda (P P^\dagger)^{p-1}) (P^\dagger)^{-1} = A(PP^\dagger)(P^\dagger)^{-1}\,,\\
  M^\dagger &:=& P^\dagger\,.
\label{trPr}
\end{eqnarray}
Then, the corresponding Jacobian can be computed as
\begin{eqnarray}
\label{SP0}
\nonumber
 \Big|\det  
\begin{bmatrix}
    \frac{dM}{dP} &  \frac{dM}{dP^\dagger}\\
    \frac{dM^\dagger}{dP}       & \frac{dM^\dagger}{dP^\dagger}
\end{bmatrix}\Big|
&=&
 \Big|\det  
\begin{bmatrix}
    \frac{dM}{dX_l} \frac{dX_l}{dP} &  \frac{dM}{dP^\dagger}\\
    \mathbf{0}       & \mathbf{1}
\end{bmatrix}\Big|\\
 &=&\Big| \det    \frac{A(Y_l) \otimes_{lr} \mathbf{1} - \mathbf{1} \otimes_{lr} A(Y_r)}
  {Y_l \otimes_{lr} \mathbf{1} - \mathbf{1} \otimes_{lr} Y_r} \Big| 
\\
\nonumber &=& \exp\Big\{\Tr_{lr}\log\Big[
  \frac{A(Y_l) \otimes_{lr} \mathbf{1} - \mathbf{1} \otimes_{lr} A(Y_r)}
  {Y_l \otimes_{lr} \mathbf{1} - \mathbf{1} \otimes_{lr} Y_r}\Big]\Big\}\, ,
\end{eqnarray}
where the symbolic matrix differentiation rule valid for analytic functions $f$ of a matrix 
\begin{equation}
\frac{\delta f(X)}{\delta X} = \frac{f(X) \otimes \mathbf{1} - \mathbf{1} \otimes f(X)}
{X \otimes \mathbf{1} - \mathbf{1} \otimes X}\, ,
\label{df}
\end{equation}
was used and the trace and tensor product acts on the Hilbert space $\cH_l \otimes \cH_r$. 
The absolute value in (\ref{SP0}) can be omitted through a perturbative regularity check.

Since it is a dummy variable, renaming $P$ as $M$ , hence $Y$ as $X$,
yields
\begin{eqnarray}
  Z(\lambda, N_l, N_r) &=& \int\, dM dM^\dagger\, \exp\{- N_r\Tr_l X_l +\LV (X_l, X_r) \} ,
\\
\LV (X_l, X_r)&=&\Tr_{lr}\log\Big[
  \frac{A(X_l) \otimes_{lr} \mathbf{1} - \mathbf{1} \otimes_{lr} A( X_r)}
  {X_l \otimes_{lr} \mathbf{1} - \mathbf{1} \otimes_{lr}  X_r}\Big]\,   .
\label{Sp1}
\end{eqnarray}
Taking into account the functional equation \eqref{AEq} one can rewrite the loop vertex action as
\begin{eqnarray}
\nonumber
\LV &= &
  \Tr_{lr}\log\Big[
    \frac{(A(X_l) +\lambda A^p(X_l)) \otimes_{lr} \mathbf{1} - \mathbf{1} \otimes_{lr} (A(X_r) +\lambda A^p(X_r))}
    {A(X_l) \otimes_{lr} \mathbf{1} - \mathbf{1} \otimes_{lr} A(X_r)}
  \Big]^{-1}  \\
\nonumber
&=& - \Tr_{lr}\log\Big[\bbone_{lr}  +\lambda
    \frac{A^p(X_l) \otimes_{lr} \mathbf{1} - \mathbf{1} \otimes_{lr} A^p(X_r)}
    {A(X_l) \otimes_{lr} \mathbf{1} - \mathbf{1} \otimes_{lr} A(X_r)}
  \Big] \\
&=&  - \Tr_{lr} \log  \Big[\bbone_{lr} +\lambda\sum_{k = 0}^{p-1} A^k(X_l) \otimes_{lr} A^{p-1-k}(X_r)
  \Big]
\,.
\label{Sp2}
\end{eqnarray}
\qed

Let us now rewrite $\LV$ in terms of $X_l$ alone. Developing the logarithm in  powers and taking the (factorized) tensor trace leads to
\bee \LV (X_l, X_r)= \sum_{n=1}^\infty \frac{(-\lambda)^n}{n} \sum_{k_1 =0}^{p-1} \cdots \sum_{k_n =0}^{p-1} [ \Tr_{l}  A^{\sum k_i}(X_l) ] [  \Tr_{r}  A^{\sum (p-k_i-1)}(X_r)  ] .
\ee
Since $A(x) = \sum_{n=1}^\infty a_n x^n $ and since $\Tr_l  X_l ^q = \Tr_r  X_r ^q $ for any $q >0$
we can rewrite everything in this  sum in terms of $X_l$ alone hence as a tensor trace on $\cH_l \otimes \cH_l$.
We have however to be careful to the fact that $\Tr_l  1_l=  N_l \ne \Tr_r  1_r =N_r$. A moment of attention therefore reveals that 
the loop vertex action $\LV$ is the sum of a ``square matrix" piece and a ``vector piece" (without any tensor product)
\bea
\LV(X_l) &=& \LV^{{\text Mat}}(X_l) + \LV^{{\text Vec}}(X_l) , \label{decompo}\\
\LV^{{\text Mat}}(X_l)&=& -  \Tr_{ll}\log\big[\bbone_{ll} +\lambda \sum_{k = 0}^{p-1} A^k(X_l) \otimes_{ll} A^{p-1-k}(X_l)\big] ,\\
 \LV^{{\text Vec}}(X_l )&=&- (N_r- N_l)
\Tr_{l}\log [\bbone_{l} +\lambda A^{p-1}(X_l) ] .
\label{Svec}
\eea
For simplicity from now on we limit ourselves  to the square case $N_l = N_r = N$, but we emphasize that our 
main result, namely \emph{uniform analyticity in $N$}, extends to the rectangular case, uniformity being in the largest
dimension $N_r $ in that case. The treatment of the additional vector piece $ \LV^{Vec}$ in \eqref{decompo} is indeed
almost trivial compared to the matrix piece, and
the corresponding details are  left to the reader.

From now on since the right space has disappeared we simply write $X$ instead of 
$X_l$, $\bbone_\otimes $ instead of $\bbone_{ll}$ etc... Our starting point rewrites in these simpler notations
\begin{eqnarray}\label{Zgood1}
Z ( \lambda, N) &=& \int\, dM dM^\dagger\, \exp\{- N \Tr \,X + \LV \} , \\
\LV(\lambda, X) &=& -  \Tr_\otimes \log\Big[\bbone_\otimes +\lambda\sum_{k = 0}^{p-1} A^k(X) \otimes A^{p-1-k}(X)
  \Big].
\label{ZAgood1}
\end{eqnarray}
Defining $\Sigma (\lambda,X) := \sum_{k = 0}^{p-1} A^k(X) \otimes A^{p-1-k}(X)$, a useful lemma is
\begin{lemma}
\bee  \frac{\partial A}{\partial X} = 
\big[\bbone_\otimes +\lambda\Sigma (\lambda,X)\big]^{-1} . \label{resoder}
\ee
\end{lemma}
\prf 
From the algebraic rule \eqref{df} and the functional equation \eqref{AEq} 
\bea  \frac{\partial  A}{\partial X} &=& \Big[\frac{(A(X)  +\lambda A^p(X)  )\otimes 1 - 1 \otimes  (A(X) +\lambda A^p(X) ) }{ A(X) \otimes 1 - 1 \otimes A (X) } \Big]^{-1}\nonumber
\\ &=& \Big[\bbone_\otimes  +\lambda \frac{ A^p(X)  \otimes 1 - 1 \otimes A^p(X)  }{ A(X) \otimes 1 - 1 \otimes A(X) } \Big]^{-1} 
\\ &=&\Big[\bbone_\otimes +\lambda \sum_{k = 0}^{p-1} A^k(X) \otimes A^{p-1-k}(X)\Big] ^{-1} = \big[\bbone_\otimes +\lambda\Sigma (\lambda,X)\big]^{-1} . \nonumber \eea
\qed

\subsection{Factorization through Holomorphic Calculus}
\label{HCalc}
We shall now establish another equivalent formula for $\LV$ \emph{factorized over left and right pieces}.
Given a holomorphic function $f$ on a domain containing the spectrum of a square 
matrix $X$, Cauchy's integral formula\footnote{It is our convention to include the 
$\frac{1}{2\text{i}\pi}$ factor of the Cauchy formula into $\oint$.} yields a convenient expression for $f(X)$,
\begin{align}
f(X)=\oint_{\Gamma} dw\frac{f(w)}{w-X},
\end{align}
provided the contour $\Gamma$ encloses the full spectrum of $X$. 

We work  with Hermitian matrices such as $X$ which have positive spectrum. Let us 
introduce a bit of notation for the contours that we shall use.

\begin{figure}[!ht]
\begin{center}
{\includegraphics[width=12cm]{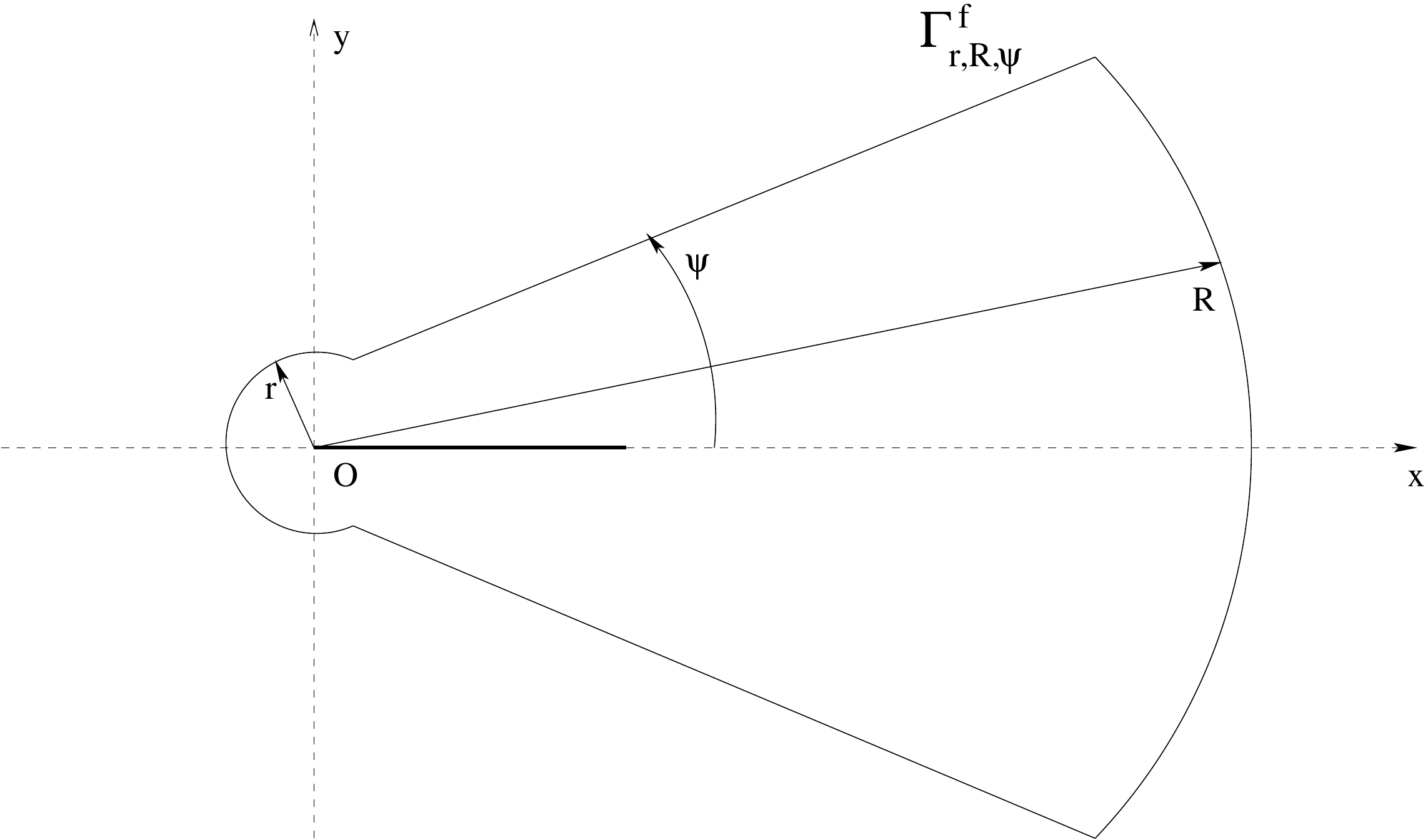}}
\end{center}
\caption{A finite keyhole contour $\Gamma^f_{r, R, \psi}$ encircling a segment 
on the real positive $Ox$ axis, shown in boldface, which includes the spectrum of $X$.}
\label{keyholeencirc0}
\end{figure}

Let's assume we have two radii $0<r<R<+\infty$ and an angle $\psi \in ] 0,\frac{\pi}{2} [$. The \emph{finite} keyhole
contour $\Gamma^f_{r, R, \psi}$ is defined as the (counterclockwise) contour in the complex plane
made of the two segments $H^\pm_{r,R ,\psi}$ joining the points $re^{\pm i \psi}$ and $Re^{\pm i \psi}$,
plus two arcs of circle namely $C_{R, \psi}$ corresponding to radius $R$ and arguments in $[- \psi, \psi]$
and $\bar C_{r, \psi}$ corresponding to radius $r$ and arguments out of $]- \psi, \psi[$, see Figure \ref{keyholeencirc0}.

Since the matrix $X$ is positive Hermitian, the condition for  holomorphic calculus is
fulfilled as soon as $R > \Vert X \Vert$. For the moment we always assume this condition to be fulfilled.

We also define the infinite keyhole contour $\Gamma^\infty_{r, \psi}$ which is the 
$R \to \infty$ limit of $\Gamma^f_{r, R, \psi}$. Of course we shall use them only
when the associated infinite contour integral is absolutely convergent.

In the following we call scalar counterparts of matrix functions by the same, but small (not capital) letters.
Thus, the function $A(\lambda, X)$ is represented by the Cauchy's formula of $a(\lambda, u)$.
We may omit the $r, R, \psi$ indices when the context is clear.

\begin{figure}[!ht]
\begin{center}
{\includegraphics[width=12cm]{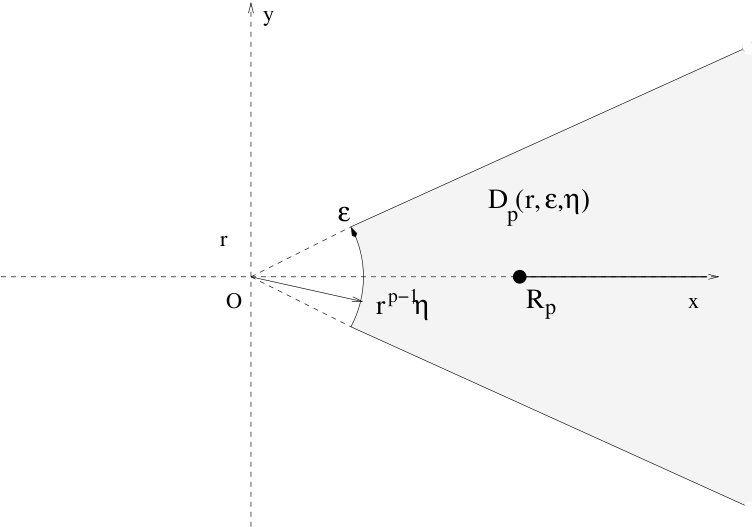}}
\end{center}
\caption{The cut-sector $D_p(r, \epsilon,\eta) $ containing the Fuss-Catalan cut starting at $R_p$
(shown in boldface).}
\label{cutsector}
\end{figure}

\begin{lemma}\label{aProp}
Suppose $\lambda \in P(\epsilon, \eta)$. For sufficiently small $r$ and $\psi$ the function $a^k(\lambda, u)$ is analytic 
in $u$ in an open neighborhood of the contour $\Gamma^f_{r, R, \psi}$ \footnote{Hereafter by this we mean the neighborhood of the contour itself and the area surrounded by the contour.} for any integer $k \in [0, p-1]$.
\end{lemma}
\prf 
Write $\lambda = \rho e^{i \theta }$ with $\rho <\eta$, $\vert \theta \vert< \pi -\epsilon$, and remember
$\Gamma^f_{r, R, \psi}= H^-_{r,R ,\psi} \cup C_{R, \psi} \cup H^+_{r,R ,\psi} \cup \bar C_{r, \psi} $. 
For $u \in H^-_{r,R ,\psi} \cup C_{R, \psi} \cup H^+_{r,R ,\psi}$ we have $\vert \arg u \, \vert \le \psi$
hence $\vert \arg u^{p-1} \vert \le (p-1)\psi $, hence if we choose $\psi < \frac{1}{2(p-1)} \epsilon $, then
 $\arg (- \lambda u^{p-1} )$ is out of the interval $[- \frac{\epsilon}{2}, \frac{\epsilon}{2} ] $. Finally if $u\in \bar C_{r, \psi} $
then $\vert - \lambda u^{p-1} \vert \le r^{p-1} \eta $. In conclusion for $u \in \Gamma^f_{r, R, \psi}$, $z =- \lambda u^{p-1}$ stays
completely out of the cut-sector 
\bee
D_p(r, \epsilon,\eta) = \{z \in {\mathbb C} , \vert z \vert \ge r^{p-1} \eta, \vert \arg z \vert \le  \frac{\epsilon}{2} \}.
\ee
Let us assume from now on that $ r^{p-1} \eta < R_p := \frac{(p-1)^{p-1}}{p^p}$.
Then this cut sector $D_p(r, \epsilon,\eta)$ fully contains the cut  of the Fuss Catalan function $T_p$
which is $[R_p,  \infty )$
\cite{Rivasseau:2017hpg,FussCatalan}. 
It follows that $T_p(- \lambda u^{p-1})$, hence also  $a^k(\lambda, u)$ for any integer $k \in [0, p-1]$, are analytic in 
$u$ in  a neighborhood of the keyhole contour 
$\Gamma^f_{r, R, \psi}$ and that the contour integrals \eqref{cauch1}-\eqref{cauchA}
are well defined.
 \qed
\medskip

We can therefore write 
\bee  A(X) =  \oint_{\Gamma} du\; a(\lambda,u)   \, \frac{1}{u-X} \label{cauch1}
\ee
where $a(\lambda, z) = z T_p(-\lambda z^{p-1})$ (see \eqref{Adef}) and the contour $\Gamma$
is a \emph{finite} keyhole contour enclosing all the spectrum of $X$.
The matrix derivative acting on a resolvent 
being easy to compute using  \eqref{df} we obtain
\bee
\frac{\partial A}{\partial X} =  \oint_{\Gamma} du\; a(\lambda, u) \frac{1}{u-X}\otimes  \frac{1}{u-X}. \label{cauchA}
\ee

Resolvent factors such as $\frac{1}{u-X}  $ are obviously non-singular on keyhole contours such as $\Gamma$ as they
have all singularities \emph{inside} by our choice of $R >\Vert X \Vert $. For safety of some formulas
below and in the next sections we shall even always 
assume $R \ge 1 + \Vert X \Vert$ so that we never even come close to a singularity of $\frac{1}{u-X}  $.
But the reader could worry about the $a$ functions in
\eqref{cauch1}-\eqref{cauchA}, in particular when $\lambda$ is complex in the pacman domain of \eqref{pacman}. This is taken care of by our next Lemma.

Combining \eqref{ZAgood1}, \eqref{resoder} and \eqref{cauchA} we get, for $\Gamma_0$ a finite keyhole contour
enclosing the spectrum of $X$
\bee \partial_\lambda \LV 
=- \sum_{k = 0}^{p-1}  \oint_{\Gamma_0} du\; a(\lambda, u) 
 \Tr_\otimes  \partial_\lambda  \big [  \frac{A^k(\lambda,X) }{u-X}  \otimes  \frac{A^{p-k-1}(\lambda,X)}{u-X}  \big] . \label{niceder0}
\ee
Now we reapply the holomorphic calculus, but in different ways\footnote{Our choices below are made in order to 
allow for the bounds of Section IV.} depending on the term chosen in the sum over $k$.

\begin{figure}[!ht]
\begin{center}
{\includegraphics[width=10cm]{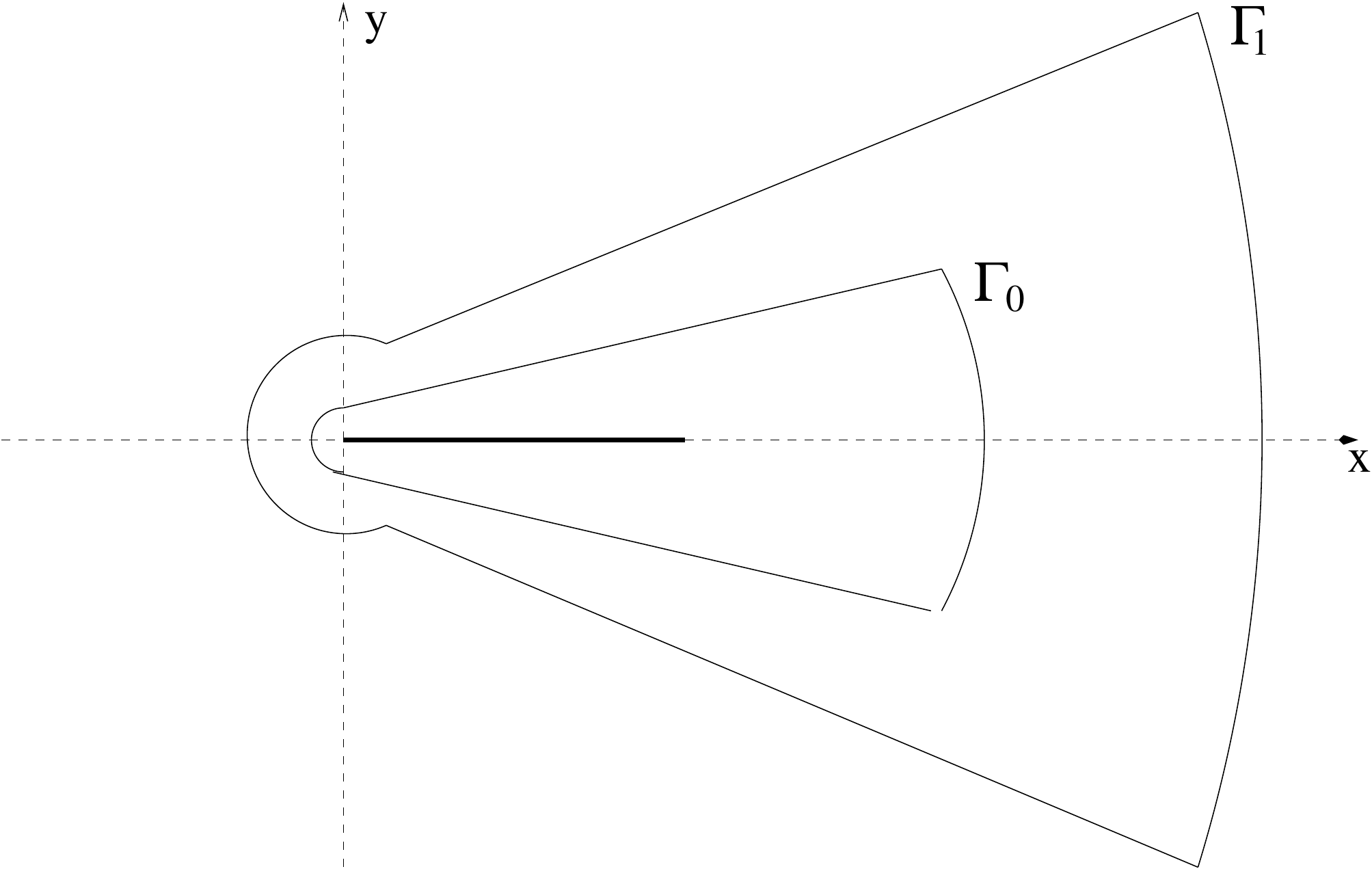}}
\end{center}
\caption{A keyhole contour $\Gamma_1$ encircling a keyhole contour $\Gamma_0$. The spectrum of $X$
lies on a real axis positive segment like the one shown in boldface.}
\label{keyholeencirc1}
\end{figure}
\begin{itemize}
\item For $k=0$, we apply the holomorphic calculus to the right $ \frac{A^{p-1}(\lambda,X)}{u-X} $ factor, with a contour $\Gamma_2$ surrounding $\Gamma_0$
for a new variable called $v_2$, and we rename $u$ and $\Gamma_0$ as $v_1$ and $\Gamma_1$ (see Figure \ref{keyholeencirc1}),

\item For $k=p-1$, we apply the holomorphic calculus to the left $ \frac{A^{p-1}(\lambda,X)}{u-X} $ factor, with a contour $\Gamma_2$ surrounding $\Gamma_0$
for a new variable called $v_2$, and we rename $u$ and $\Gamma_0$ as $v_1$ and $\Gamma_1$; we obtain a contribution identical to the previous case.

\item In all other cases, hence for $1\le k \le p-2$ we apply the holomorphic calculus both to left and right factors in the tensor product,
with two variables $v_1$ and $v_2$ and two equal contours $\Gamma_1$ and $\Gamma_2$ enclosing enclose the contour $\Gamma_0$ .

\end{itemize}

In this way  defining the ``loop resolvent"
\bee \cR (v_1, v_2, X) :=   \label{resolv1}
 \Big[\Tr   \frac{1}{v_1-X} \Big]  \Big[\Tr   \frac{1}{v_2-X} \Big] 
\ee
we obtain
\bea  \frac{\partial \LV}{\partial \lambda} &=&- \oint_{\Gamma_1} dv_1 \oint_{\Gamma_2} dv_2  \Big\{ \oint_{\Gamma_0} du\; a(\lambda, u) \sum_{k = 1}^{p-2} 
\frac{\partial_\lambda [\lambda a^k(\lambda,v_1) a^{p-k-1}(\lambda,v_2)]}{(v_1 - u)(v_2 - u)} \, \nonumber \\
&& \qquad +\; 2 a(\lambda, v_1) \frac{ \partial_\lambda \big[ \lambda a^{p-1}(\lambda,v_2) \big]  }{v_1 - v_2}  \Big\}\cR (v_1, v_2, X) .
 \label{niceder1}
\eea

%


Therefore defining the weights 
\bee
\phi (\lambda, u, v_1, v_2) := -\sum_{k = 1}^{p-2}  
\frac{1 }{v_1 - u}  \frac{1}{v_2 - u}a(\lambda, u) \partial_\lambda \big[ \lambda a^k(\lambda,v_1) a^{p-k-1}(\lambda,v_2) \big]  \label{defphi}
\ee
\bee
\psi (\lambda, v_1, v_2) :=  -  \frac{2}{v_1 - v_2}  a(\lambda, v_1)   \partial_\lambda \big[ \lambda a^{p-1}(\lambda,v_2) \big]  
\ee
we have, provided $\Gamma_0$, $\Gamma_1$ and $\Gamma_2$ are finite keyholes contours all enclosing
$[0, \Vert X \Vert ]$ and $\Gamma_1$ and $\Gamma_2$ enclose $\Gamma_0$:
\begin{lemma}\label{lemmafactor}
\bea
\LV (\lambda, X)&=&  \int_0^{\lambda} dt  \oint_{\Gamma_1} dv_1   \oint_{\Gamma_2}dv_2  \Big\{  \oint_{\Gamma_0} du
\ \phi (t, u, v_1, v_2) \nonumber \\
&& \qquad + \;\; \psi (t, v_1, v_2)  \Big\} \;  \cR (v_1, v_2, X) .
 \label{goodfactor}
\eea
\end{lemma}
\proof Simply remark that $\LV \vert_{\lambda=0} =0$ and apply first order Taylor formula, using \eqref{niceder1}.
\qed
\medskip

The two traces in $\cR$ can be thought either as the two sides of a single \emph{ribbon loop} or as
two independent ordinary \emph{loops} (hence the name loop vertex representation). 
Remark indeed that these two loops are \emph{factorized} in $\cR$. They
are coupled only through the scalar factors (the $u$ contour integral for the $\phi$ term
or the $(v_1-v_2)^{-1}$ factor for the $\psi$ term). The condition on the contours
 $\Gamma^f_{r_j, \psi_j, R_j}$ for $j= 0,1, 2$, can be written $0< r_0 < \min (r_1, r_2)$; $0< \psi_0 < \min (\psi_1, \psi_2) \le   \max (\psi_1, \psi_2) < \pi - \epsilon $
 and $\Vert X \Vert  + 1  \le R_0 < \min (R_1, R_2)$.  

The nice property of this representation is that it does not break the symmetry between the two factors in the tensor product. Beware that the three $\Gamma$ contours in \eqref{goodfactor} have to be finite ones $\Gamma^f_{r_j, \psi_j, R_j}$,
hence \emph{not universal in $X$}, since they depend on $X$ through the condition that $R_0$
must be strictly bigger than $\Vert X \Vert$.  

A careful study using the bounds of Section \ref{LVEbound} reveals that the finiteness of these three contours, hence their $X$-dependence,
cannot be removed because the integral  \eqref{goodfactor} is not absolutely convergent as $R \to \infty$
(this is linked to the fact that $\LV$ is \emph{not uniformly bounded in $X$} but grows logarithmically at large $X$).
Fortunately this slightly annoying feature will \emph{fully disappear
in the LVE formulas below}, because these formulas  do not use $\LV$ but derivatives of $\LV$
with respect to the field $M$ or $M^\dagger$. These derivatives are uniformly bounded. Therefore
contours of the LVE amplitudes can be taken as \emph{infinite} keyholes $\Gamma^\infty_{r, \psi}$
which are then completely independent of $X$.

\section{The Loop Vertex Expansion}

To generate a convergent loop vertex expansion \cite{Rivasseau:2007fr, Rivasseau:2017hpg},
we start by expanding the exponential of the effective action $\LV(X)$ in \eqref{Zgood1} into the Taylor series
\bea 
Z(\lambda, N) &=& \sum_{n = 0}^\infty \frac{1}{n!} 
  \int\, dM dM^\dagger \, \exp\{- N\Tr X \}\, \LV^n  \,.
\label{LVE1}
\eea
The next step is to introduce replicas and to replace (for the term of order $n$)
the integral over the single $N\times N$ complex matrix $M$ by an integral over an $n$-tuple of such $N\times N$
matrices $M_i,  1\leq i \leq n$. 
The Gaussian part of the integral is replaced  by a normalized
Gaussian measure $d\mu_C$ with a degenerate covariance $C_{ij} = N^{-1} \; \; \forall i, j $. Recall that for any real positive
symmetric matrix $C_{ij}$ one has
\begin{equation}
\int d\mu_C M^\dagger_{i|ab} M_{j|cd} = C_{ij} \delta_{ad}\delta_{bc},
\end{equation} 
where $M_{i|ab}$ denote the matrix element in the row $a$ and column $b$ of the matrix $M_i$.
That Gaussian integral with a degenerate covariance is indeed equivalent to 
a single Gaussian integral, say over $M_1$ 
times a product of $n-1$ Dirac distributions
$\delta(M_1-M_2)\cdots \delta(M_{n-1}-M_{n})$. From the perturbative point of view, this degenerate
covariance produces all the edges in a Feynman graph expansion that connect the various vertices together.
The partition function can be written as
\bea 
Z(\lambda, N) &=& \sum_{n = 0}^\infty \frac{1}{n!} 
  \int\, d\mu_C\,\prod_{i = 1}^n\, \LV(M_i)\,.
\label{LVE2}
\eea
The generating functional can be represented as a sum over the set $\gF_n$
of forests $\cF$ on $n$ labeled vertices\footnote{Oriented forests simply distinguish edges $(i,j)$
and $(j,i)$ so have edges with arrows.It allows to distinguish below between operators 
$\frac{\partial}{\partial M^\dagger_i}\frac{\partial}{\partial M_j^{\textcolor{white}\dagger}}$ and $\frac{\partial}{\partial M^\dagger_j}\frac{\partial}{\partial M_i^{\textcolor{white}\dagger}}$.}
by applying the BKAR formula \cite{BK,AR1} to \eqref{LVE2}.
We start by replacing the covariance $C_{ij} = N^{-1}$ by $C_{ij}(x) = N^{-1}x_{ij}$ ($x_{ij} = x_{ji}$) evaluated at $x_{ij} = 1$
for $i \neq j$ and $C_{ii}(x) = N^{-1} \; \forall i$. Then the Taylor BKAR formula yields
\bea 
Z(\lambda,N)&=&\sum_{n=0}^\infty\frac{1}{n!}\;\;\sum_{\cF\in \gF_n}\;
\int dw_\cF\  \partial_\cF  {\cal Z}_n\;  \Big|_{x_{ij} = x_{ij}^\cF (w)}
\label{LVE3}
\eea
where
\bea 
\int dw_\cF &:= & \prod_{(i, j) \in \cF} \int_0^1dw_{ij} \; ,  \quad
\partial_\cF  := \prod_{(i, j) \in \cF} \frac{\partial}{\partial x_{ij}} \; ,\\
{\cal Z}_n &:=& \int\, d\mu_{C(x)} \,\prod_{i = 1}^n\, \LV(M_i)
\label{LVEzn}
\\
x_{ij}^\cF &=& \left\{ 
\begin{array}{c}
\hspace{-1.4cm}\text{inf}_{(k,l)\in P^\cF_{i\leftrightarrow j}} w_{kl}~~~~~~~~~~~ \text{if}~ P^\cF_{i\leftrightarrow j}~ \text{exists}\,, \\ 
0~~~~~~~~~~~~~~~~~~~~~~~~~~~~\text{if}~ P^\cF_{i\leftrightarrow j}~ 
\text{does not exist}\, .
\end{array}
\right.
\label{LVE4a}
\eea
In this formula $w_{ij}$ is the weakening parameter of the edge $(i,j)$ of the forest, 
and $P^\cF_{i\leftrightarrow j}$ is the unique path in $\cF$ joining $i$ and $j$ when it exists.

Substituting the contour integral representation \eqref{goodfactor} for each $\cS (M_i)$  factor
in \eqref{LVE2}, we rewrite \eqref{LVEzn} as
\bee  {\cal Z}_n =  \int\, d\mu_{C(x)}\, \int  \{dt du dv\} \Phi_n   \cR_n
\ee
where $\cR_n$ stands for the product of all resolvents
\bee   \cR_n := \prod_{i = 1}^n  \cR_i ( v^i_1 , v^i_2, X_i) , \label{cln}
\ee 
and the symbol $\int  \{dt du dv\} \Phi_n  \cR_n$ stands for
\bea \int  \{dt du dv \} \Phi_n   \cR_n  &=& \prod_{i = 1}^n \Bigg[  \int_0^\lambda dt^i \oint_{\Gamma^i_1} dv^i_{1} \oint_{\Gamma^i_2}  dv^i_{2} 
\Big\{\oint_{\Gamma^i_0}  du^{i}  \phi (t^i, u^i , v^i_1 , v^i_2)  \nonumber\\
&+&\psi (t^i, v^i_1 , v^i_2)  \Big\}  \cR_i\Bigg] \label{clnphi}  
\eea
where  the contours areas specified in the previous section. We put most of the time 
in what follows the 
replica index $i$ in upper position but beware not to confuse it with a power. 
Since  Gaussian integration can be represented as a differentiation
\begin{equation}
  \int\, d\mu_{C(x)} \, f(M) = \Big[e^{\frac{1}{N}\sum_{i,j} x_{ij} \Tr\big[\frac{\partial}{\partial M^\dagger_i}\frac{\partial}{\partial M_j^{\textcolor{white}\dagger}}\big]}
f(M)\Big]\Big|_{M_i = 0}\,.
\label{Idiff}
\end{equation}
Then, the differentiation with respect to $x_{ij}$ in (\ref{LVE5}) results in
\begin{equation}
\frac{\partial}{\partial x_{ij}}\Big(\int\, d\mu_{C(x)} \, f(M)\Big) 
= \frac{1}{N} \int\, d\mu_{C(x)}  \Tr\big[\frac{\partial}{\partial M^\dagger_i}\frac{\partial}{\partial M_j^{\textcolor{white}\dagger}}\big] f(M)\,.
\end{equation}
The operator $\Tr\big[\frac{\partial}{\partial M^\dagger_i}\frac{\partial}{\partial M_j^{\textcolor{white}\dagger}}\big]$ acts on two distinct
loop vertices ($i$ and $j$) and connects them by an oriented edge.
Introducing the notation 
\bee
\partial_\cF^M = 
 \prod_{(i, j) \in \cF} \Tr\big[\frac{\partial}{\partial M^\dagger_i}\frac{\partial}{\partial M_j^{\textcolor{white}\dagger}}\big] 
\ee
we can commute all functional derivatives in $\partial_\cF$ with all contour integrals, using the argument 
of Section \ref{HCalc} that the contours are far from the singularities of the integrand. We can then also 
commute the functional integral and the contour integration. This results in
\bee 
Z(\lambda, N) = \sum_{n = 0}^\infty \frac{1}{n!}\,\sum_{\cF\in \gF_n} N^{-\vert \cF \vert}  \int dw_\cF 
\int \{dt du dv\} \Phi_n  \int d\mu_{C(x)} \partial_\cF^M   \cR_n
 \Big|_{x_{ij} = x_{ij}^\cF (w)} .
  \label{LVE4}
\ee

As usual, since the right hand side of \eqref{LVE4}
is now factorized over the connected components of the forest $\cF$, which are spanning trees, 
its logarithm, which selects only the connected parts, is expressed by \emph{exactly the same formula} but summed
over \emph{trees}. For a tree on $n$ vertices $\vert \cT \vert  = n-1$. Taking into account
the $N^{-2}$ factor in the normalization of $F$ in \eqref{Fsq} we obtain the 
expansion of the free energy as (remark the sum which starts now at $n=1$ instead of $n=0$)
\bee 
F(\lambda , N)= \sum_{n = 1}^\infty \frac{1}{n!}\,\sum_{\cT\in \gT_n} A_\cT\label{LVE6}
\ee
\bee
A_\cT :=  N^{-n-1}   \int dw_\cT 
\int \{dt du dv \} \Phi_n  \int d\mu_{C(x)}  \partial_\cT^M   \cR_n
\Big|_{x_{ij} = x_{ij}^\cT (w)},
\label{LVE5}
\ee
where $\gT_n$ is the set of oriented spanning trees over $n\ge 1$ labeled vertices.

Our main result is
\begin{theorem} \label{main} For any $\epsilon>0$ there exists $\eta$ small enough such that
the expansion \eqref{LVE6} is absolutely convergent and defines an analytic function of $\lambda$,
uniformly bounded in $N$, in the ``pacman domain" 
\bee
P(\epsilon, \eta) := \{0< |\lambda |<\eta, \vert \arg \lambda \vert < \pi - \epsilon\},
\ee
a domain which is \emph{uniform in $N$}. Here absolutely convergent and uniformly bounded in $N$ means
that for fixed $\epsilon$ and $\eta$ as above there exists a constant $K$ independent of $N$ such that  
for $\lambda \in P(\epsilon, \eta)$
\bee   \sum_{n = 1}^\infty \frac{1}{n!}\,\sum_{\cT\in \gT_n}  \vert A_\cT \vert \le K < \infty . \label{unifobou}
\ee

\end{theorem}
Absolute convergence  would be of course wrong for the usual expansion
of $F$ into connected Feynman graphs. Moreover the difficult part of the theorem is the \emph{uniformity}  in $N$ of the domain $P(\epsilon, \eta) $ and of the bound \eqref{unifobou}. 
Indeed the fact that $F$ is analytic and in fact Borel-Le Roy summable of order $p-1$ (in the Nevanlinna-Sokal sense of \cite{Sok,Lionni:2016ush}) but in a domain which \emph{shrinks} with $N$ as $N \to \infty$
is already known, see eg \cite{Lionni:2016ush}.

The next subsection is devoted to compute explicitly $ \partial_\cT^M   \cR_n$, and Section IV is devoted to bounds which 
prove this Theorem.

\subsection{Derivatives of the action}
\label{secder}

We need now to compute $ \partial_\cT^M   \cR_n$. This will be relatively easy since $\cR_n$ is a product of resolvents
of the $\frac{1}{u-X}$ type. 
Since trees have arbitrary coordination numbers we need a formula
for the action on a vertex factor $\cR^i $ of a certain number $r^i=q^i+ \bar q^i$ of
derivatives, $q^i$ of them of the $\frac{\partial }{\partial M_i }$ type and $\bar q_i $ of the 
$\frac{\partial }{\partial M_i^\dagger} $ type.

Let us fix a given loop vertex and forget for a moment the index $i$. We need
to develop a formula for the action of a differentiation operator 
$\frac{\partial^r }{\partial M_1 \cdots \partial M_q  \partial M^\dagger_1  \cdots \partial M^\dagger_{\bar q} } $ on $\cR = \Tr \frac{1}{v_1-X}  \otimes  \Tr \frac{1}{v_2-X}  $.

To perform this computation we first want to know on which of the two traces (also simply called ``loops") of a loop vertex the differentiations act.
Therefore we add to any oriented tree $\cT$ of order $n$ a collection of  $2(n-1)$ indices $s_e$. Each such index takes value in $\{1,2\}$, 
and specifies at each end $e$ of an edge of the tree whether the field derivative for this end hits the $ \Tr \frac{1}{v_1-X}$ loop or the  
$ \Tr \frac{1}{v_2-X}$ loop. 
There are therefore exactly $2^{2(n-1)}$ such decorated 
oriented trees for any oriented tree. Unless otherwise specified
in the rest of the paper we simply use the word ``tree" for an oriented decorated tree with these additional $\{s \}$ data.
Similarly the set $\cT_n$ from now on means the set of oriented \emph{decorated} trees at order $n$. 

Knowing the decorated tree $\cT$, at each vertex we know how to decompose the number of differentiations
acting on it according to a sum over the two loops of the number of differentiations on that loop, as $q = q_1 + q_2$, $\bar q= \bar q_1 + \bar q_2$.
Hence we have the simpler problem to compute the differentiation operator  
$\frac{\partial^r }{\partial M_1 \cdots \partial M_{q}  \partial M^\dagger_1  \cdots \partial M^\dagger_{\bar q}} $ on 
a \emph{single} loop $\Tr \frac{1}{v-X} $.

We shall use the symbol $\sqcup$ to indicate the place where the indices of the derivatives act\footnote{The symbol $\sqcup$
instead of $\otimes$ will hopefully convey the fact that these derivatives are half propagators for the LVE. The edges of the LVEs always glue two $\sqcup$ symbols together.}.
For instance we shall write 
\bee  \frac{\partial }{\partial X}  \frac{1}{v-X} =    \frac{1}{v-X} \sqcup \frac{1}{v-X} .
\ee
To warm up let us compute explicitly some derivatives (writing $\partial_M$ for  $\frac{\partial }{\partial M}$) \:
\begin{eqnarray}
\nonumber
\partial_M \Tr \frac{1}{v-X}  &= &\Big[\Tr\frac{1}{v-X} \sqcup M^\dagger  
 \frac{1}{v-X} \Big] 
\label{eqder1}\\
\partial_{M^\dagger}  \Tr \frac{1}{v-X} &= &\Big[\Tr\frac{1}{v-X} M \sqcup \frac{1}{v-X}  \Big]  .
\label{eqder2}
\end{eqnarray}
Induction is clear: $r=q+\bar q$ derivatives create insertions of $\sqcup M^\dagger  $ and of 
$M \sqcup$ factors in all possible cyclically distinct  ways but they can 
also create double insertions noted $ \sqcup \sqcup $ when a $M^\dagger  $
or $M$ numerator is hit by a derivative. For instance at second order
we have:
\begin{eqnarray}
\nonumber
\partial_M^{\textcolor{white}\dagger}  \partial_{M^\dagger}  \Tr \frac{1}{v-X}  
&= &\Tr\Big[\frac{1}{v-X} M \sqcup \frac{1}{v-X} \sqcup M^\dagger  
 \frac{1}{v-X} \Big] 
 \nonumber\\
 &+ &\Tr\Big[\frac{1}{v-X} \sqcup M^\dagger  
 \frac{1}{v-X} M \sqcup \frac{1}{v-X} \Big]  
 \nonumber\\
 &+ &\Tr\Big[\frac{1}{v-X} \sqcup \bbone \sqcup
 \frac{1}{v-X} \Big]  .
 \label{eqder3}
\end{eqnarray}
Remark the last term in which the second derivative hits the numerator created by the first. 
Since $X =M M^\dagger$ the outcome for a $q$-th order partial derivative, 
is a bit difficult to write,
but the combinatorics is quite inessential for our future analyticity bounds.
The Fa\`a di Bruno formula allows to write this outcome as as sum over a set  $\Pi^{q, \bar q}_r$ of Fa\`a di Bruno  terms 
each with prefactor 1:
\bee \frac{\partial^r }{\partial M_1 \cdots \partial M_q  \partial M^\dagger_1  \cdots \partial M^\dagger_{\bar q} }\Tr  \frac{1}{v-X} 
= \sum_{\pi \in \Pi^{q, \bar q}_r} \;\Tr\Big[O^\pi_0 \sqcup O^\pi_1 \sqcup \cdots \sqcup O^\pi_r \Big]. \label{faasum}
\ee
In the sum \eqref{faasum} there are exactly $r$  symbols $\sqcup$, separating $r+1$ corner operators $O^\pi_c$.
These corner operators can be of four different types, either resolvents $\frac{1}{v-X}$, $M$-resolvents $\frac{1}{v-X} M $, $M^\dagger$-resolvents 
$M^\dagger \frac{1}{v-X}  $,
or the identity operator $\bbone$. We call $r_\pi$, $r^M_\pi$, $r^{M^\dagger}_\pi$ and $i_\pi$ the number of corresponding operators in $\pi$.
We shall need only the following facts.
\begin{lemma} We have
\bee \vert \Pi^{q, \bar q}_r \vert \le   2^r  r!, \quad
r_\pi   =   1  +  i_\pi    ,  \quad   r^M_\pi  +r^{M^\dagger}_\pi = r - 2 i_\pi .
\ee \label{faacomb}
\end{lemma}
\proof Easy by induction, since at order $r$ for each new derivative we have to hit any of the $r$ 
$O_c$ operators of order $r-1$ (hence the $r!$ factor), and eventually if that operator is an $M$-resolvent
or $M^\dagger$-resolvent of the right type we can decide with a further factor 2 to hit either the resolvent or
the $M$ (or $M^\dagger)$ factor. The rest of the Lemma is trivial. \qed

Applying \eqref{faasum} at each of the two loops of each loop vertex, we get for any decorated tree $\cT$
\bee  \partial_\cT^M   \cR_n = \prod_{i=1}^n \Big\{ \prod_{j= 1}^2 \Big[ \sum_{\pi_j^i \in \Pi_{r^i_j}^{q^i_j, \bar q^i_j  }  } 
\Tr  \big( O^{\pi_j^i}_0 \sqcup O^{\pi_j^i}_1 \sqcup \cdots \sqcup O^{\pi_j^i}_{r^i_j} \big) \Big]  \Big\}\label{faafullsum}
\ee
where the indices of the previous \eqref{faasum} are simply all decomposed into indices for each loop $j=1, 2$
of each loop vertex $i = 1 ,\cdots , n $.

\begin{figure}[!ht]
\begin{center}
{\includegraphics[width=12cm]{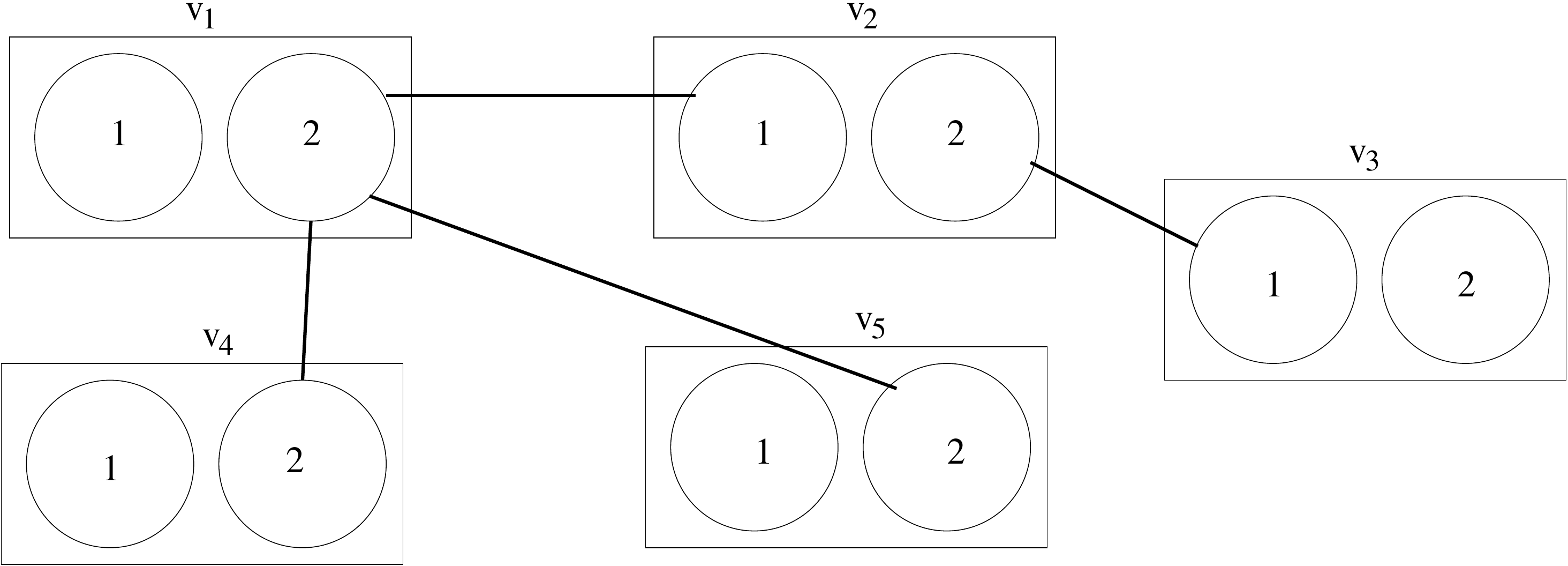}}
\end{center}
\caption{A tree of $n-1$ lines on $n$ loop vertices (depicted as rectangular boxes, hence here $n=5$) 
defines a forest of $n+1$ connected components or cycles $\cC$ on the $2n$ elementary loops, since
each vertex contains exactly two loops. To each such cycle
corresponds a  trace of a given product of operators in the LVE.}
\label{forest}
\end{figure}

We need now to understand the gluing of the $\sqcup$ symbols. 
Knowing the decoration of the tree, that is the $2(n-1)$ indices $s_e$, we know exactly for which edge of the decorated
tree which loops it connects. In other words the \emph{decorated}
tree $\cT_n$ defines a particular \emph{forest} on the $2n$ loops of the $n$ loop vertices (see Figure \ref{forest}). This forest having $n-1$ 
edges must therefore have exactly
$n+1$ connected components, each of which is a tree but on the $2n$ loops. We call these trees the \emph{cycles} $\cC$ of the tree,
since as trees, they have a single face.

Now a moment of attention reveals that if we fix a particular choice $\{ \pi_j^i \}$ in \eqref{faafullsum}
expansion obtained by the action of $ \partial_\cT^M $ on $\cR_n$ the $\sqcup$ symbols
since they are summed with indices forced to coincide along the edges of the tree simply
glue the $2n$ traces of \eqref{faafullsum} into $n+1$ traces, one for each cycle $\cC$ of the decorated tree $\cT$.
This is the fundamental feature of the LVE \cite{Rivasseau:2007fr}.
Each trace acts on the product of all corners operators $O_c$ cyclically ordered in the way obtained by  turning around the cycle $\cC$. Hence 
we obtain, with hopefully transparent notations,
\bee \partial_\cT^M   \cR_n = \prod_{i=1}^n \Big\{ \prod_{j= 1}^2 \Big[  \sum_{\pi_j^i \in \Pi_{r^i_j}^{q^i_j, \bar q^i_j  }  }  \Big] \Big\} \prod_{\cC}
 \Bigl[ \Tr  \prod_{c  \; \circlearrowleft \; \cC} O_c \Big] .  \label{faafullsumcycl}
\ee
We now bound the associated tree amplitudes of the LVE.

\section{LVE amplitude bound}
\label{LVEbound}

The beauty of the LVE method is that the associated amplitudes can be bounded by a convergent geometric series
\emph{uniformly} in $w$, $M$ and $N$. From now on let us suppose first that $n\ge 2$, hence
$\cT$ is not the trivial tree $\cT_\emptyset$ with one vertex and no edge. 
To bound the amplitude of this trivial tree $\cT_\emptyset$ is much easier but requires, 
as usual in any LVE, a particular treatment given in Section \ref{trivialtree}. We perform first the functional integral bound, then the contour integral bound.
For that we rewrite \eqref{LVE5} as
\bea A_\cT (\lambda,N )&=&
\int \{dt du dv \} \Phi_n F_\cT ( \lambda, N, u,v)   \\
F_\cT ( N, v) &:=& N^{-n-1}   \int dw_\cT  \int d\mu_{C(x)}  
\prod_{i=1}^n \Big\{ \prod_{j= 1}^2 \Big[ \sum_{\pi_j^i \in \Pi_{r^i_j}^{q^i_j, \bar q^i_j  }  } \Big] \Big\}\nonumber  \\
&&\prod_{\cC}
 \Bigl[ \Tr  \prod_{c  \; \circlearrowleft \; \cC} O_c \Big] 
\Big|_{x_{ij} = x_{ij}^\cT (w)}.
\eea
and bound first the functional integral  $F_\cT$.

\subsection{Functional Integral Bound}
Starting from \eqref{faafullsumcycl}
we simply bound every trace by the dimension of the space, which is $N$, times the product of the norms of all operators
along that cycle. This is the same strategy than in \cite{Rivasseau:2007fr}. 
Since there are exactly $n+1$ traces, 
the factors $N$ \emph{exactly cancel}, all operator norms now 
commute as they are scalars, and taking into account Lemma \ref{faacomb} we are left with
 \bee   \vert  F_\cT ( N, v)  \vert  \le 2^{2n-2} \prod_{i=1}^n r_i ! \int dw_\cT  \int d\mu_{C(x)}  \sup_{\pi}
\prod_{c} \Bigl[  \Vert O_c \Vert \Big]
 \Big|_{x_{ij} = x_{ij}^\cT (w)} \label{funcbou1}
 \ee
Using that $\sup\{  \Vert M \Vert , \Vert M^\dagger \Vert  \}\le \Vert X\Vert^{1/2}$, 
it is easy to now bound resolvent factors, for $v$'s on these keyhole contours, by
\bea
\Vert \frac{1}{v^i_j - X^i} \Vert  &\le& K  (1 + \vert v^i_j  \vert)^{-1} , \\
\Vert \frac{1}{v^i_j - X^i} M^i  \Vert  &\le& K (1 + \vert v^i_j  \vert)^{-1/2}, \\
\Vert {M^i}^\dagger \frac{1}{v^i_j - X^i} \Vert  &\le& K (1 + \vert v^i_j  \vert)^{-1/2} .
\eea
where $K$ denotes a generic constant which depends on the contour parameters $r$ and $\psi$. Plugging into
\eqref{funcbou1} we can use again Lemma \ref{faacomb} to prove that we get exactly a decay factor $ (1 + \vert v^i_j  \vert)^{-(1+ r^i_j/2)} $
for each of the $2n$ loops. The corresponding bound being \emph{uniform} in all $\pi, \{w\}, \{M\}$, and since the integrals
$\int dw_\cT  \int d\mu_{C(x)} $ are \emph{normalized}, we get
\bee   \vert  F_\cT ( N, v)  \vert  \le 2^{2n-2} K^n\prod_{i=1}^n \Big\{ r_i ! 
\prod_{j= 1}^2  (1 + \vert v^i_j  \vert)^{-(1+ r^i_j/2)} \Big\}\label{funcbou2}
\ee
Recall that with our notations, $r_i =  r^i_1 +  r^i_2$.

\subsection{Contour Integral Bound}

We insert now the bound \eqref{funcbou2} in the contour integral $ \int \{dt du dv \} $, of course taking absolute values in the integrand, since we took an absolute value
for $\vert  F_\cT (N, v) \vert $. Our integration contours being complex we use the shortened notation 
$\vert \int_\Gamma \vert f(z) \vert dz \vert $ 
to mean $\int \vert f(z) \vert \vert \frac{dz}{dx} \vert dx $ where $x$ is a real variable
parametrizing the contour $\Gamma$. With these shortened notations, and absorbing the $2^{2n-2} $ factor
by changing the value of $K$, we get

\bee \vert A_\cT \vert \le  K^n \Big\vert  \int \{dt du dv\} \vert \Phi_n  \vert \prod_{i=1}^n \Big\{ r_i ! \prod_{j= 1}^2  (1 + \vert v^i_j  \vert)^{-(1+ r^i_j/2)} \Big\} \Big\vert .  \label{contou1}
\ee
Remark that this bound is now \emph{factorized} over the loop vertices, since $\Phi_n$ is factorized, see \eqref{clnphi}. Hence we shall now 
fix again a vertex of index $ i$ and we omit to write the $i$ superscript for a while for the reader's comfort. 

Remark that  $r_1 + r_2 = r >0$ (since each $r_i$ in \eqref{contou1} is strictly positive, because $\cT$ is not the trivial tree). 
Since \eqref{contou1} is a decreasing function of $r_1$, $r_2$,
we need only to bound the worst cases, namely $r_1=1$, $r_2=0$ or $r_1=0$, $r_2=1$. Since $\phi$ is symmetric in $v_1$, $v_2$, but not $\psi$, we end up with three 
different integrals to bound:
\bee \cI_1=\Big\vert  \int dt du dv_1 d v_2 \vert \phi (t,u,v_1,v_2)  \vert   (1 + \vert v_1 \vert)^{-3/2}  (1 + \vert v_2 \vert)^{-1} 
\Big\vert , \label{contou3}
\ee
\bee \cI_2=\Big\vert  \int dt dv_1 d v_2 \vert \psi (t,v_1,v_2)  \vert   (1 + \vert v_1 \vert)^{-3/2}  (1 + \vert v_2 \vert)^{-1} \Big\vert ,\label{contoupsi1}
\ee
\bee \cI_3=\Big\vert  \int dt dv_1 d v_2 \vert \psi (t,v_1,v_2)  \vert   (1 + \vert v_1 \vert)^{-1}  (1 + \vert v_2 \vert)^{-3/2} \Big\vert .\label{contoupsi2} 
\ee

Returning to the definition \eqref{defphi} of $\phi$  we have first to compute the derivative 
\bea
\partial_t\big[  t a^k(t,v_1)a^{p-k-1}(t, v_2)  \big]  &=&  \Big[   a^k(t,v_1)a^{p-k-1}(t, v_2) \label{defphi1} \\
&+& t 
[k \partial_t a(t,v_1)  a^{k-1}(t,v_1)a^{p-k-1}(t, v_2) \nonumber \\
&+& (p-k-1) \partial_t a(t,v_2)  a^{k}(t,v_1)a^{p-k-2}(t, v_2)] .\Big]\nonumber 
\eea
Therefore we need a bound on the factors $a(t, v)$ and $\partial_t a(t, v)$ for $v$ on a keyhole contour.
Recalling Lemma III.1 in  \cite{Rivasseau:2017hpg}, for $z$  in the complement of $D_p(r, \epsilon,\eta) $ we 
have
\bea \vert  T_p(z) \vert 
&\le& \frac{K}{(1 + \vert z\vert)^{1/p}} ,\label{decay1}
\\ \vert \frac{d}{dz} T_p(z) \vert 
&\le& \frac{K}{(1 + \vert z\vert)^{1 + \frac{1}{p}} }. \label{decay2}
\eea 
Since $a(t,v) = v T_p (-t v^{p-1})$ we
find that for $v\in \Gamma$ 
\bea \vert a (t, v) \vert 
&\le&   \vert v \vert \frac{K}{(1 +\vert t \vert \vert v   \vert^{p-1})^{1/p}}  ,\label{decaya1}
\\
 \vert \partial_t a(t,v)  \vert &\le&\vert  v\vert ^{p } \frac{K}{(1 +\vert t \vert \vert v   \vert^{p-1})^{1 + \frac{1}{p}} }   .
\label{decaya2}
\eea 
Defining
\bee  \tilde A_k (t,v) := \frac{ \vert v \vert^{k} }{ ( 1 +\vert t \vert \vert v   \vert^{p-1})^{\frac{k}{p}}  }  ,
\label{AkBound}
\ee
since $\frac{\vert t \vert \vert v \vert^{p-1} }{ ( 1 +\vert t \vert \vert v   \vert^{p-1}) } \le 1$  we find
\bee \vert \frac{\partial }{\partial t}\big[  t a^k(t,v_1)a^{p-k-1}(t, v_2)  \big] \vert \le 
K  \tilde A _k (t,v_1)\tilde A_{p-k-1} (t,v_2)  .
\label{supdt}
\ee
We now insert this bound in the contour integral $\cI_1$ and find
\bea \cI_1 &\le& \Big\vert  \int dt du dv_1 d v_2 \frac{ \tilde A _1 (t,u)}{\vert u-v_1\vert \vert u-v_2 \vert}  (1 + \vert v_1 \vert)^{-3/2}  (1 + \vert v_2 \vert)^{-1}\nonumber \\
&&\sum_{1 \le k \le p-2 } \tilde A _k (t,v_1) \tilde A_{p-k-1} (t,v_2)    \Big\vert . \label{contou4a}
\eea
We need now to use a bit of convex analysis. On our contours we have $\vert u-v \vert \ge K (1 + \vert u\vert )$ and
$\vert u-v \vert \ge K (1 +\vert v\vert) $, hence for any $(\alpha_1, \alpha_2) \in [0,1]^2$ we have
\bee    \frac{1}{\vert u-v_1\vert \vert u-v_2 \vert}  \le (1 + \vert u\vert )^{-2 + \alpha_1 + \alpha_2}   (1 + \vert v_1\vert )^{- \alpha_1}
 (1 + \vert v_2\vert )^{- \alpha_2} .
\label{uvuv}
\ee
Furthermore for any $0 \le \beta \le 1$ we have
\bee \tilde A_k (t,v) \le \vert v \vert^{k - \beta k (p-1)/p    } \vert t \vert^{-\beta k/p}  .
\label{Abeta}
\ee
Therefore for any choice of the five numbers $(\alpha_1, \alpha_2 , \beta, \beta_1, \beta_2) \in [0,1]^5$ we have
\bea \cI_1 &\le&\sum_{1 \le k \le p-2 }   \Big\vert  \int_0^\lambda dt  \vert t \vert^{- \frac{1}{p} (\beta  + k\beta_1  +(p-k-1) \beta_2 ) } 
\oint_{\Gamma_0} du \frac{\vert u \vert^{1 - \beta (p-1)/p    }}{(1 + \vert u\vert )^{2 - \alpha_1 - \alpha_2} }
 \nonumber \\
&& \oint_{\Gamma_1} dv_1 \frac{\vert v_1 \vert^{k(1 - \beta_1 (p-1)/p )}  }{(1 + \vert v_1 \vert)^{3/2 +\alpha_1}}   
\oint_{\Gamma_2}  d v_2 \frac{ \vert v_2 \vert^{(p-k-1)(1 - \beta_2 (p-1)/p )} }{
 (1 + \vert v_2 \vert)^{1+ \alpha_2}  }  \Big\vert .\label{contou4}
\eea
We choose $\beta= 1-\epsilon_p$, $\beta_1 = \beta_2 = 1$ and $\alpha_2 = \frac{p-k-1}{p} + \epsilon_p$ 
and distinguish two cases. If $1 \le k < \frac{p}{2}$ we choose $\alpha_1 =0$.
Substituting in \eqref{contou4} we find after some trivial manipulations such as $1+ \vert u \vert \ge \vert u \vert $ we get
\bea \cI _1&\le&\sum_{1 \le k \le p-2 }    \Big\vert  \int_0^\lambda dt    \vert t \vert^{- 1 + \frac{\epsilon_p}{p}}
\oint_{\Gamma_0} du \frac{\vert u \vert^{\frac{1+ (p-1) \epsilon_p}{p} }}{(1 + \vert u\vert )^{1 + \frac{k+1}{p} - \epsilon_p   } }
 \nonumber \\
&& \oint_{\Gamma_1} dv_1  (1 + \vert v_1 \vert)^{-\frac{3}{2}+\frac{k}{p}}  
\oint_{\Gamma_2}  d v_2 
 (1 + \vert v_2 \vert)^{-1- \epsilon_p}  \Big\vert . \label{contou5}
\eea
The $v_1$ and $v_2$ contour integrals are now absolutely convergent and bounded by ($p$-dependent) 
constants. Since $k \ge 1$, the $u$ integral is also convergent for small $\epsilon_p$, for instance
$\epsilon_p = \frac{1}{2p}$, since for that choice its integrand then behaves, in the worst case $k=1$, as $u^{-1 -  \frac{1}{2p^2} }$.

If $\frac{p}{2} \le k \le p-2$ we choose $\alpha_1 =\frac{k}{p} - \frac{1}{2}  + \epsilon_p $ and get
\bea \cI _1&\le&\sum_{1 \le k \le p-2 }    \Big\vert  \int_0^\lambda dt    \vert t \vert^{- 1 + \frac{\epsilon_p}{p}}
\oint_{\Gamma_0} du \frac{\vert u \vert^{\frac{1+ (p-1) \epsilon_p}{p} }}{(1 + \vert u\vert )^{\frac{3}{2} +  \frac{1}{p} -2 \epsilon_p } }
 \nonumber \\
&& \oint_{\Gamma_1} dv_1  (1 + \vert v_1 \vert)^{-1-\epsilon_p}  
\oint_{\Gamma_2}  d v_2 
 (1 + \vert v_2 \vert)^{-1- \epsilon_p }  \Big\vert . \label{contou5a}
\eea
The three contour integrals are now absolutely convergent and bounded by  ($p$-dependent) constants
for instance if we choose $\epsilon_p = \frac{1}{4p}$ (since $p \ge 2$).

We conclude that
\bee \cI_1 \le   K \vert \lambda \vert^{\kappa_p} \label{contou6a}
\ee
for $\kappa_p := \frac{1}{4p^2} >0$ (we do not try to optimize this number).

The bounds on $\cI_2$ and $\cI_3$ are  simpler. Inserting \eqref{supdt} in \eqref{contoupsi1}-\eqref{contoupsi2} we find
\bee \cI_2\le K \Big\vert  \int dt dv_1 d v_2 \tilde  A _1 (t,v_1) \tilde A_{p-1} (t,v_2)   \frac{1}{\vert v_1-v_2 \vert}  (1 + \vert v_1 \vert)^{-3/2}  (1 + \vert v_2 \vert)^{-1} \Big\vert ,\label{contoupsi3}
\ee
\bee \cI_3\le K \Big\vert  \int dt dv_1 d v_2 \tilde  A_{p-1} (t,v_1) \tilde A _1 (t,v_2)    \frac{1}{\vert v_1-v_2 \vert}  (1 + \vert v_1 \vert)^{-3/2}  (1 + \vert v_2 \vert)^{-1} \Big\vert . \label{contoupsi4}
\ee
On our contours we have 
$\vert v_1-v_2 \vert \ge K (1 + \vert v_1\vert )$ and
$\vert v_1-v_2 \vert \ge K (1 +\vert v_2\vert) $, hence for any $\alpha \in [0,1]$ we have
\bee   
\frac{1}{\vert v_1-v_2\vert}  \le (1 + \vert v_1\vert )^{-\alpha} (1 + \vert v_2\vert )^{- (1-\alpha)} .
\label{v1v2}
\ee
Therefore using again \eqref{Abeta} for any choice of the three numbers $(\alpha ,\beta_1, \beta_2) \in [0,1]^3$ we have
\bea \cI_2 &\le& K   \Big\vert  \int_0^\lambda dt   \vert t \vert^{- \frac{1}{p} (\beta_1  +(p-1) \beta_2 ) } 
\oint_{\Gamma_1} dv_1 \frac{\vert v_1 \vert^{1 - \beta_1 (p-1)/p }  }{(1 + \vert v_1 \vert)^{3/2 +\alpha}}   \\
&&
\oint_{\Gamma_2}  d v_2 \frac{ \vert v_2 \vert^{(p-1)(1 - \beta_2 (p-1)/p )} }{
 (1 + \vert v_2 \vert)^{2- \alpha}  }  \Big\vert  . \label{contoupsi5}
\eea
We can choose $\beta_1 = \frac{1}{2}$, $\beta_2 = 1$, $\alpha = \frac{3}{4p}$ and we get
\bea \cI_2 &\le& K   \Big\vert  \int_0^\lambda dt   \vert t \vert^{- 1 + \frac{1}{2p}} \oint_{\Gamma_1} dv_1 \frac{1 }{(1 + \vert v_1 \vert)^{1+ \frac{1}{4p}}  }
\oint_{\Gamma_2} dv_2 \frac{1 }{(1 + \vert v_2 \vert)^{1+ \frac{1}{4p}} } \nonumber\\
&\le & K \vert \lambda \vert^{\frac{1}{2p}} . \label{contoupsi6}
\eea
Similarly we have for any choice of the three numbers $(\alpha ,\beta_1, \beta_2) \in [0,1]^3$
\bea \cI_3 &\le& K   \Big\vert  \int_0^\lambda dt   \vert t \vert^{- \frac{1}{p} ((p-1)\beta_1  + \beta_2 ) } 
\oint_{\Gamma_1} dv_1 \frac{\vert v_1 \vert^{(p-1)(1 - \beta_1 (p-1)/p )}  }{(1 + \vert v_1 \vert)^{3/2 +\alpha}}   \\
&&
\oint_{\Gamma_2}  d v_2 \frac{ \vert v_2 \vert^{1 - \beta_2 (p-1)/p } }{
 (1 + \vert v_2 \vert)^{2- \alpha}  }  \Big\vert .  \label{contoupsi7}
\eea
We can choose $\beta_1 = 1$, $\beta_2 = \frac{1}{2}$, $\alpha = \frac{1}{2}- \frac{3}{4p}$ and we get the same outcome
\bea \cI_3 &\le& K   \Big\vert  \int_0^\lambda dt   \vert t \vert^{- 1 + \frac{1}{2p}} \oint_{\Gamma_1} dv_1 \frac{1 }{(1 + \vert v_1 \vert)^{1+ \frac{1}{4p}}  }
\oint_{\Gamma_2} dv_2 \frac{1 }{(1 + \vert v_2 \vert)^{1+ \frac{1}{4p}} } \nonumber\\
&\le & K \vert \lambda \vert^{\frac{1}{2p}} . \label{contoupsi8}
\eea

%
%

We can gather our results in the following lemma:
\begin{lemma} \label{mainamp}
For any $\epsilon>0$, there exists $\eta_{\epsilon}>0$ and a constant  $K>0$  such that for any 
tree $\cT$ with $n$ vertices the amplitude $ A_{\cT}(\lambda,N)$ is analytic 
in $\lambda$ in the pacman domain $P(\epsilon, \eta_\epsilon)$
and satisfies in that domain to the \emph{uniform} bound in $N$
\begin{align} \label{mainbou}
 | A_\cT (\lambda,N )| \le K^n \vert \lambda \vert ^{\kappa_p n} \prod_{i=1}^n r_i ! 
\end{align}
where $r_i \ge 1$ is the coordination of the tree $\cT$ at vertex $i$. 
\end{lemma}
\proof We simply put together \eqref{contou6a} and \eqref{contoupsi6}-\eqref{contoupsi8}. \qed

By Cayley's theorem, the number of labeled trees with coordination $r_i$ on $n$ vertices is
$\frac{(n-2)!}{ \prod_{i=1}^n (r_i-1)! }$. Orientation and decoration add to the bound  an inessential factor $2^{3(n-1)}$. Furthermore
$\prod r_i \le 2^{\sum_i r_i} = 2^{2 (n-1)}$.
Remembering the symmetry factor $\frac{1}{n!} $ in \eqref{LVE6},
Theorem \ref{main} follows now easily from Lemma \ref{mainamp}. Remark that the right hand side in \eqref{mainbou} is independent of $N$, 
so that our results hold uniformly in $N$.

\subsection{The trivial $n=1$ tree}
\label{trivialtree}

To bound the trivial tree amplitude $A_{\cT_\emptyset}$ with a single vertex, hence $n=1$, namely
\bee A_{\cT_\emptyset}= \int\, d\mu \ \LV (\lambda, X) \label{trivtree}
\ee
obviously does not require replicas but requires a little additional step namely integration
by parts of one field. We return to \eqref{Zgood1}-\eqref{ZAgood1} 
but now single out one of the $X= M M^\dagger$ factors in the $\Sigma (\lambda , X)$ numerator
and \emph{do not write it} through the holomorphic calculus technique.
Hence we insert in \eqref{trivtree} the slightly different representation
\begin{lemma}\label{lemmafactor1}
\bea
\LV (\lambda, X)&=&  \int_0^{\lambda} dt  \oint_{\Gamma_1} dv_1   \oint_{\Gamma_2}dv_2 \ \Big\{  \oint_{\Gamma_0} du
 \Big[\phi_1 (t, u, v_1, v_2)\nonumber \\
&+&  \ \psi_1 (t, v_1, v_2)  \Big] \Big\} [ \Tr  \frac{1}{v_1 - X} ][\Tr  \frac{X}{v_2 - X} ]
 \label{goodfactortriv}
\eea
with the new contour weights
\bea
&\phi_1 (t, u, v_1, v_2) :=  - \sum_{k = 1}^{p-2}  \frac{ a(t, u)}{(u -  v_1)(u -  v_2)} 
\frac{\partial }{\partial t}\big[  a^{k}(t,v_1)  t T_p (t, v_2) a^{p-k-2}(t, v_2)  \big] \nonumber \\
&\psi_1 (t, v_1, v_2) := 
 -  \frac{2}{v_1 - v_2}  a(t, v_1)  
\frac{\partial }{\partial t}\big[  t T_p (t, v_2) a^{p-2}(t, v_2)  \big] \label{defpsi}
\eea
where $T_p (t, v) = T_p (-t v^{p-1})$.
\end{lemma}
\proof Exactly similar to the one of Lemma \ref{lemmafactor}. \qed

We then perform a single step of integration by parts 
on the $M$ factor in the numerator $X$ of  $\Tr  \frac{X}{v_2 - X}$ in
\eqref{goodfactortriv} with respect to the functional measure. 
It gives
\bea && A_{\cT_\emptyset}= N^{-3} \int\, d\mu  \int_0^{\lambda} dt  \oint_{\Gamma_1} dv_1   \oint_{\Gamma_2}dv_2   \Big\{  \oint_{\Gamma_0} du
\ \Big[\phi_1 (t, u, v_1, v_2)  \nonumber  \\ 
&& + \psi_1 (t, v_1, v_2)  \Big] \Big\} 
\Big\{\sum_{abc} \frac{\partial}{\partial M^\dagger_{ba} } \Big[(M^\dagger \frac{1 }{v_2 - X})_{ba}( \frac{1}{v_1 - X})_{cc} \Big]\Big\}.
\label{goodfactortriv1}
\eea
Beware indeed that the indices of the $\partial_{M^\dagger}$ operator dual to $M$ 
in the Gaussian integration by parts are inserted in the $v_2$ trace but as a differential operator it can also act on the  $v_1 $ trace. 
The $\partial_{M^\dagger} $ factor is then computed as in section \ref{secder}. Following  the same steps
than in the previous section and using the gain of one $v_1$ or $v_2$ numerators on $\phi_1$ or $\psi_1$ 
compared to $\phi$ and $\psi$ we get (more easily!) a bound for $\lambda \in P(\epsilon, \eta)$
\bee \vert A_{\cT_\emptyset} \vert \le  \vert \lambda \vert ^{\kappa_p }  K . \label{trivtree1}
\ee
Observe that this bound is uniform in $N$ because we have either one or three traces (depending whether
the $\partial_{M^\dagger} $ acts on the  $ \Tr  \frac{1}{v_1 - X} $
or the $\Tr  \frac{ M^\dagger }{v_2 - X} $) and three $\frac{1} {N}$ factors.

\section{Discussion and Conclusion}
Pushing further the functional integration over the replica fields creates additional loops on the tree $\cT$. If the loop is
\emph{planar}, we get a factor of $1/N$ (new edge) and a factor of $N$ (new face), so that the scaling is left unchanged. 
But if the loop adds a non planar edge (an edge that connects distinct faces), then the scaling is reduced by a factor $1/N^{2}$.

In this way we can build a constructive version of the topological expansion up to any fixed genus $g$ similar to 
the one of \cite{Gurau:2014lua}. Also cumulants could presumably be studied 
exactly as in \cite{Gurau:2014lua}. This is left to the reader in order
to keep this paper more readable.

It seems now clear that the full reparametrization invariance of Feynman's functional integral has
not been fully exploited yet. Of course applying the idea of the LVR to
ordinary quantum field theory leads to non-local interactions. Nevertheless
non-local interactions are nowadays more studied than in the past because of the quantum
gravity problematic \cite{Rivasseau:2016zco}. Time has perhaps come to look at them with a fresh eye.
We intend in any case to explore the consequence of the very general idea of the LVR 
for tensor models \cite{Guraubook} and for ordinary field theories in future publications.

\renewcommand{\theequation}{\thesection.\arabic{equation}}
\appendix

\section{Effective action via the selective Gaussian integration}
\label{partinteg}
We give now a second proof of Theorem \ref{theoremlvr} for the case of square matrices performing the selective Gaussian integration.
This integration can be implemented by employing the formula \eqref{Idiff}, representing integration as differentiation:
\begin{eqnarray}
\nonumber
  Z =
  e^{\Tr[\frac{1}{N}\partial_M^{\textcolor{white}\dagger} \partial_{M^\dagger}]} e^{N\Tr[-\lambda (M M^\dagger)^{p - 1} M M^\dagger]}\Big|_{M^\dagger = 0, M = 0} = \\
  e^{\Tr\big[\frac{1}{N}\partial_M \partial_{A}\big]} 
  \Big(
    e^{\frac{1}{N}\Tr[\partial_M \partial_{B}]} 
    e^{N \Tr[-\lambda (M A)^{p-1} M B]}\Big|_{B = 0}
  \Big)\Big|_{A = 0, M = 0}
  \,,
\label{ZC3}
\end{eqnarray}
where, for instance,
  $\Tr[\partial_M \partial_{M^\dagger}] = 
  \sum_{a,b} \frac{\partial}{\partial M_{ab}^{\textcolor{white}\dagger} } \frac{\partial}{\partial M_{ba}^\dagger}$.
Renaming matrix $A$ back as $M^\dagger$, we obtain
\begin{eqnarray}
  Z = \int\, dM\, \exp\{- N \Tr[M M^\dagger] + \LV(M)\}
\end{eqnarray}
with the effective action given by
\begin{eqnarray}
  \LV = \log
  \Big(
    e^{\frac{1}{N}\Tr[\partial_M \partial_{B}]} 
    e^{N \Tr[-\lambda (M M^\dagger)^{p-1} M B]}\Big|_{B = 0}
  \Big)\,.
\end{eqnarray}
Within the framework of the perturbation theory, the exponent of $\LV$ can be transformed as
\begin{eqnarray}
\nonumber
  e^{\LV} &=& e^{\frac{1}{N}\Tr[\partial_C \partial_{B}]} 
  e^{N\Tr[-\lambda ((M + C) M^\dagger)^{p-1} (M + C) B]}\Big|_{B = 0, C = 0}\\
  &=& \int\, dC\,dB\, e^{-N \Tr[C B] + N \Tr[-\lambda ((M + C) M^\dagger)^{p-1} (M + C) B]}\,.
\end{eqnarray}
Then, the matrix $B$ can be interpreted as a Lagrange multiplier and
\begin{eqnarray}
\nonumber
  e^{\LV} &=& \int\, dC\, \delta(C + \lambda ((M + C) M^\dagger)^{p-1} (M + C))\\
\nonumber
  &=& \int\, dC\, \delta(C - C_0) \Big|\det\frac{\delta(C + \lambda ((M + C) M^\dagger)^{p-1} (M + C))}{\delta C}\Big|^{-1}
  \\
\nonumber
  &=& \exp\{-\Tr_\otimes[\log\big|\mathbf{1}^{LR}_\otimes + 
  \lambda\sum_{k=0}^{p-1}[(C_0 + M) M^\dagger]^k \otimes [(C_0 + M) M^\dagger]^{p - 1 - k}\big|]\}
  \,,\\
\label{eSp}
\end{eqnarray}
where the cyclicity of the trace of the formal power series determined by the logarithm was taken into account
to obtain corresponding ordering of matrices $M^\dagger$.
The matrix $C_0$ in (\ref{eSp}) is a solution of the equation
\begin{equation}
  C_0 + \lambda ((M + C_0) M^\dagger)^{p-1} (M + C_0) = 0
\end{equation}
and it is given by
\begin{equation}
  C_0 = MM^\dagger T_p(-\lambda (M M^\dagger)^{p-1})(M^\dagger)^{-1} - M\,,
\label{c0}
\end{equation}
where $T_p(z)$ is the generation function of the Fuss-Catalan numbers.
Combining (\ref{c0}) and (\ref{eSp})  we arrive at \eqref{ZAgood1}.

\section{Integral representation of the Fuss-Catalan functions}
\label{FussCatSec}
The Fuss-Catalan numbers of order $p-1$ (with $p$ being the degree in the functional equation \eqref{FCEq}), are defined by
\begin{equation}
  FC_{p-1}(n) := \frac{1}{(p-1)n + 1}    
  \left(\begin{array}{c}
      (p-1)n + n\\
      n
   \end{array}\right)\,.
\end{equation}
They can be represented as moments 
\begin{equation}
  FC_{p-1}(n) = \int_0^{1/R_{p}} dx\, x^n P_{p-1}(x)
\end{equation}
of the distribution
\begin{equation}
  P_{p-1}(x) = {\cal M}^{-1}[FC_{p-1}(\sigma), x]\,,
\end{equation}
where ${\cal M}^{-1}$ is the inverse \emph{Mellin transform} \cite{Penson2011}. 
The direct \emph{Mellin transform} of a function $f(x)$ and its inverse are defined by
\begin{eqnarray}
  f^*(\sigma) := {\cal M}[f(x), \sigma] = \int_0^\infty dx\, x^{\sigma-1} f(x)
\end{eqnarray}
and
\begin{eqnarray}
  f(x) := {\cal M}^{-1}[f^*(\sigma), x] = \frac{1}{2 \pi i} \int_{c - i\infty}^{c + i\infty}d\sigma\, x^{-\sigma} f^*(\sigma)
\end{eqnarray}
with complex $\sigma$. According to \cite{Penson2011}, properties of the \emph{Mellin transform} of a convolution lead to
\begin{equation}
  P_{p-1}(x) > 0\,,~~~\text{for}~x \in ]0 , 1/R_p[\,.
\label{posP}
\end{equation}
Then, the Fuss-Catalan generation function can be expressed as
\begin{equation}
  T_p(z) = \int_0^{1/R_{p}} \frac{dx}{1 - z x} P_{p-1}(x)\,.
\label{irep}
\end{equation}
The formula \eqref{irep} provides an analytic continuation for the Fuss-Catalan series
to the cut plane $\mathbb{C}^{cut}_{p}:= \mathbb{C} - [R_p, + \infty]$.
In particular, it follows from \eqref{posP} and \eqref{irep} that
\begin{equation}
  T_p(z) > 0 ~~ \text{for}~~z < 0\,.
\label{pos}
\end{equation}
The latter positivity property is crucial for proving the non-perturbative correctness of the effective action 
\eqref{ZAgood} (see Appendix \ref{CorrVCh}).

\section{Non-perturbative correctness of the effective action}
\label{CorrVCh}
In this section we justify the effective action \eqref{ZAgood} beyond the formal power series level.
\begin{lemma}
  For all $\lambda > 0$ the transformation \eqref{trPr} is bijective
  and corresponding \emph{Jacobian} \eqref{SP0} is positive.
\end{lemma}
\proof
The inverse transformation to \eqref{trPr} is given by
\begin{eqnarray}
  P = (M M^\dagger + \lambda (MM^\dagger)^p) (M^\dagger)^{-1}\,,~~~P^\dagger = M^\dagger\,.
\end{eqnarray}
Consequently, \eqref{trPr} is a bijection.

Choosing the basis, where $X = MM^\dagger$ is diagonal, denoting the X eigenvalues by $s_i$, we rewrite \eqref{ZAgood} as
\begin{eqnarray}
\LV(X) = - \sum_{i, j}\log\Big[1 + \lambda \sum_{k = 0}^{p-1} a^k(\lambda, s_i) a^{p-1-k}(\lambda, s_j) \Big].
\end{eqnarray}
The eigenvalues $s_i > 0$, consequently, according to \eqref{pos} $a^k(\lambda, s_i) > 0$ and 
for $\lambda > 0$ the function $\LV(X)$ is real and the \emph{Jacobian} \eqref{SP0} is positive.
\qed

The lemma above validates the change of variables leading to the effective action \eqref{ZAgood}
for $\lambda > 0$. Consequently, according to uniqueness of analytic continuation,
the action \eqref{ZAgood} defines the \emph{same} non-perturbative 
partition function and free energy as the initial one \eqref{Zsq}, \eqref{Fsq}. Using the LVR we
can prove its analyticity  in $\lambda$ in the $N$-independent
``pacman domain'' \eqref{pacman}, something which was not known using the initial
representation.
Remember that in \cite{Lionni:2016ush} Borel Le-Roy summability (of order $p-1$) of the free energy 
was established for the complex matrix models, but in a \emph{domain which shrinked} as $N \to\infty$.
Our main theorem now establishes analyticity
of the \emph{same non-perturbative free energy function}, but in a domain which \emph{no longer shrinks as $N \to\infty$}.


\section{Relationship with Perturbation Theory}

Compared to the quartic LVE case \cite{Rivasseau:2013ova},
it is a bit more difficult to describe the subset of (pieces of) Feynman graphs that the LVR developed in this paper associates  to a given
LVE tree. The resolvent $\cR$ in \eqref{resolv1} is very simple but the link
to Feynman graphs is somewhat hidden in the complicated $\phi$ and $\psi$ functions of \eqref{goodfactor}.

In this last Appendix we explain this relationship. First, in the following subsection \ref{informal}, 
we do it in an informal way, giving an intuitive understanding without additional formulas.
Then, in the subsection \ref{l2}, we compare (up to the order $\lambda^2$) the standard perturbation theory obtained with the initial polynomial action and the one obtained
with the logarithmic effective action. In the subsection \ref{LVESPT}, we match our loop vertex expansion with the terms of the standard perturbation theory for the case of
the quartic interaction, $p=2$.


\subsection{Informal explanation}
\label{informal}
\begin{figure}[!ht]
\begin{center}
{\includegraphics[width=7cm]{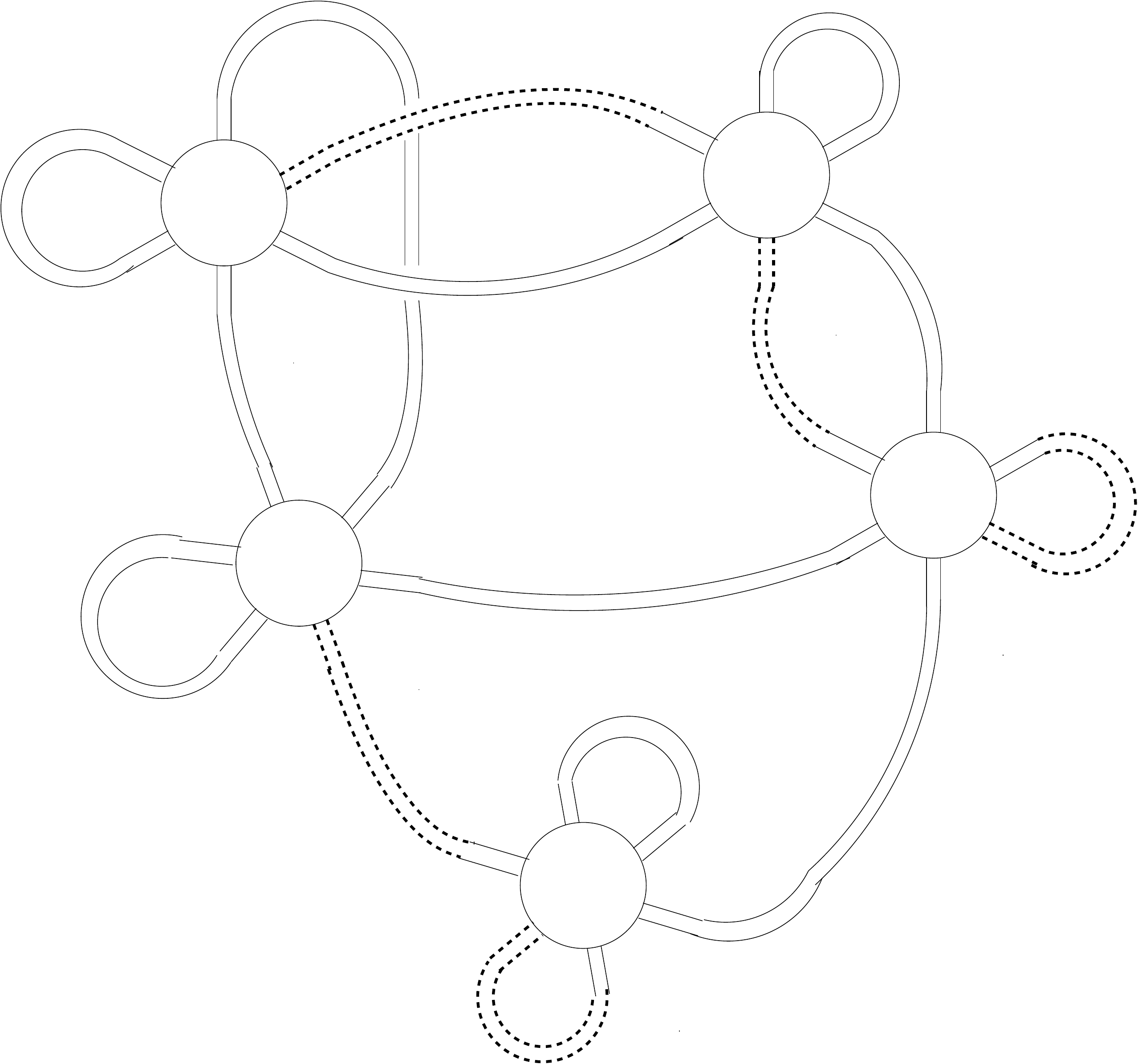}}
\end{center}
\caption{A Feynman graph of the $\Tr (M^\dagger M )^3$ theory. All five vertices are 6-valent. One $M^\dagger$
field per vertex leads to a dotted line (for better lisibility we showed as dotted the hooked point plus a large fraction of the propagator) 
defines in this case two connected components, namely two loop vertices, 
each of which has exactly one red loop.}
\label{feynman1}
\end{figure}

Let us return to
the partial integration point of view of Appendix \ref{partinteg}.
First, we explain how to visualize  the LVR vertices associated to a Feynman graph.
Taking an ordinary connected Feynman graph, we draw,
at every vertex of the graph, one selected half-edge (say corresponding to an $M^\dagger$ variable) as a dotted half-line. The set of edges which are so dotted then defines a subset of connected components, each of which has a \emph{single loop}. They are the LVR \emph{vertices} associated
to this graph (see Figure \ref{feynman1}). 

\begin{figure}[!ht]
\begin{center}
{\includegraphics[width=7cm]{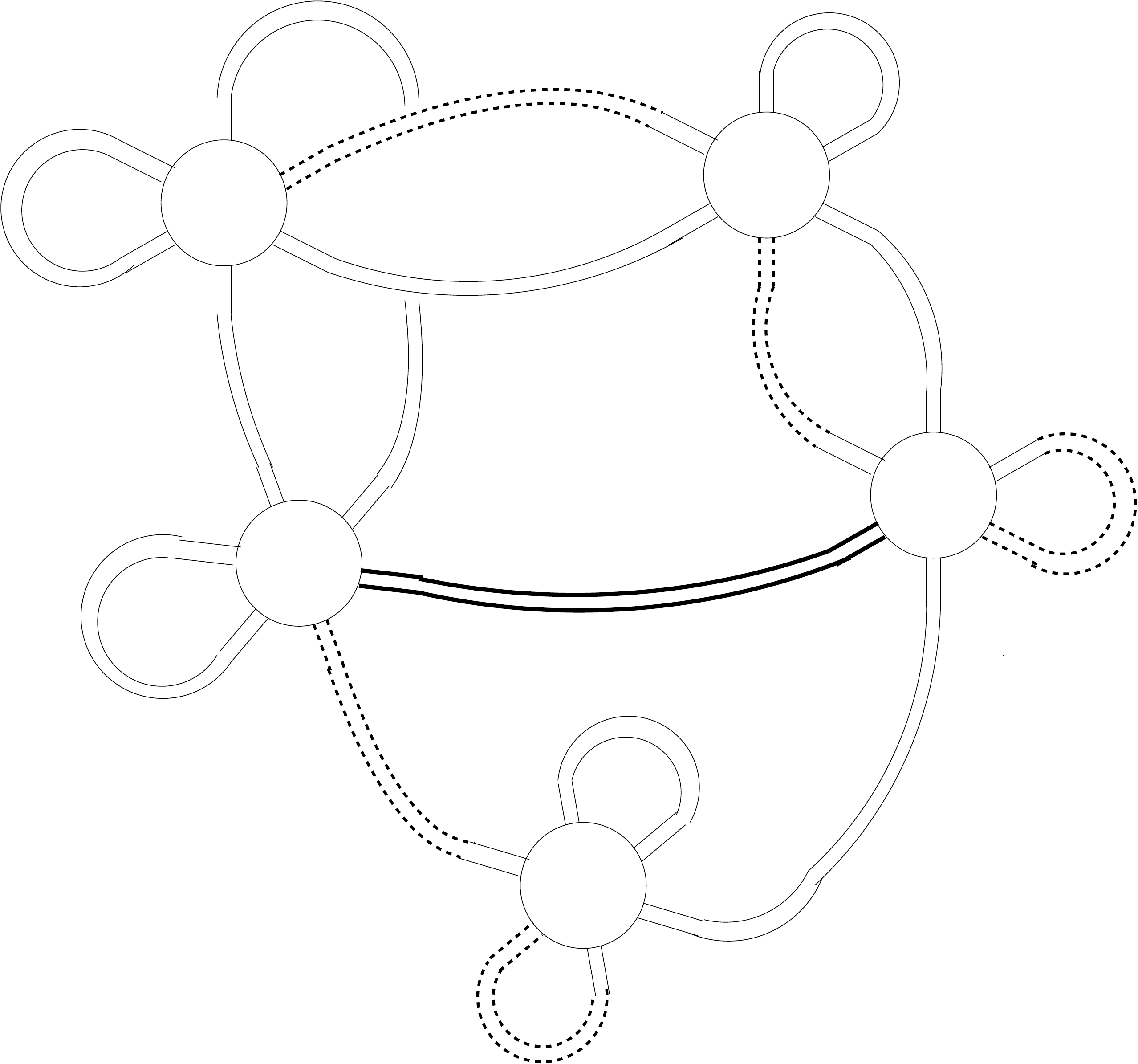}}
\end{center}
\caption{Adding a tree line in boldface between the two loop vertices
gives one of the Feynman graph contributions to the LVE tree made of two loop vertices
joined by a single edge.}
\label{feynman2}
\end{figure}

Selecting a spanning tree between these vertices through the BKAR formula, is like dividing each
Feynman graph built around these LVR vertices into as many pieces as there are 
of \emph{spanning trees} between them. Each piece is then attributed to the corresponding LVE tree
(see Figure \ref{feynman2}).

\begin{figure}[!ht]
\begin{center}
{\includegraphics[width=8cm]{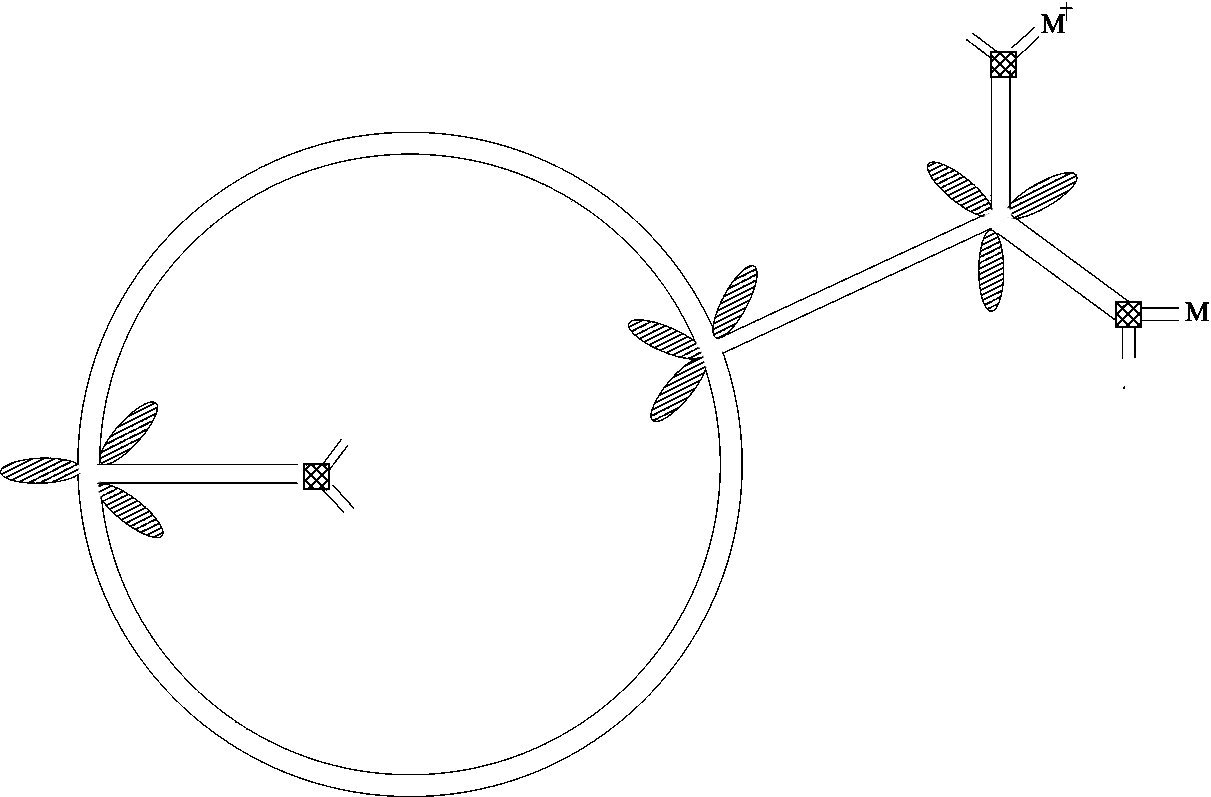}}
\end{center}
\caption{A loop vertex of the theory, bearing 4 derivatives, hence four sources $\sqcup$. 
We chose $p=5$, hence all vertices are 6-valent. Hatched ellipses represent $A_p$ insertions, ribbon edges represent resolvents
(there are five such resolvents in this graph) and squares represent derived leaves 
which can be of three different types $M \sqcup$, $\sqcup M^\dagger$ or $\sqcup \bbone \sqcup$. In the case pictured, we have three squares because
two derivatives acted on the same $M^\dagger$ factor.}
\label{matrixlve2}
\end{figure}

Conversely if we start from a given LVE tree and want to picture the whole set 
of (pieces of) Feynman graphs that it sums, we have to return to \eqref{ZAgood1} and introduce
a symbol, such as a hatched ellipse, to picture the sum of all p-ary trees in the 
generating $A_p$ function. A loop vertex of the theory can be then pictured as in Figure \ref{matrixlve2},
where the cilium and each derived leaf bear a factor $\sqcup$, each edge bears a (tensor) resolvent $R$
and each ordinary leaf bears a factor $A_p$.
\begin{figure}[!ht]
\begin{center}
{\includegraphics[width=11cm]{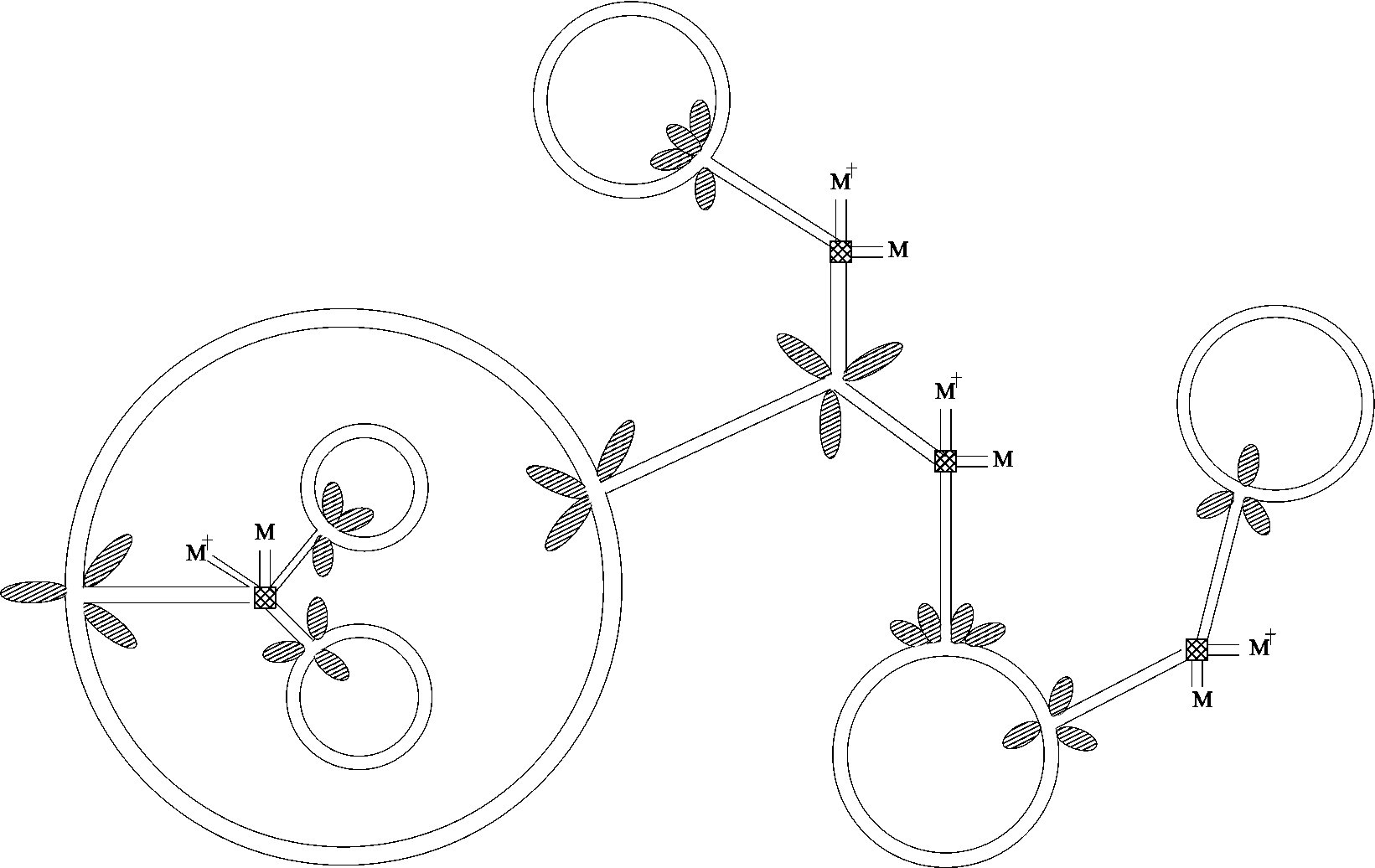}}
\end{center}
\caption{A tree of the loop vertex expansion. It is made of  six loop vertices, 
joined by four edges each bearing a square, which corresponds to the gluing of two $\sqcup$
of the previous picture, and the attentive reader can find seven traces in the drawing.}
\label{matrixlve1}
\end{figure}

Similarly a LVE tree is obtained by gluing $n$ such loop vertices through along $n-1$ pairs of 
glued $\sqcup$ factors, see Figure \ref{matrixlve1}.
Beyond the tree, additional cycles between the loop vertices can of course exist but they are hidden in the 
functional integral $ \int dw_\cF  \int d\mu_{C(x)} $  in \eqref{LVE4}. 

\clearpage
\subsection{Equivalence at order $\lambda^{2}$ for the effective action for general $p$}
\label{l2}

The effective action \eqref{ZAgood1} reads
\begin{eqnarray}
S &=& -\Tr_\otimes \log\Big[\bbone_\otimes +\lambda\sum_{k = 0}^{p-1} A^k(X) \otimes A^{p-1-k}(X)\Big]\nonumber\\
&=& \text{Tr}\log\Big[\frac{A(X) \otimes 1 - 1 \otimes A(X)}{X\otimes 1-1\otimes X}\Big]\,,
\end{eqnarray}
remember $X = MM^\dagger$.
Substituting 
\begin{align}
A(\lambda,X)&=XT(-\lambda X^{p-1})=X-\lambda X^{p}+p\lambda^{2}X^{2p-1}+O(\lambda^{3})\,,
\end{align}
yields
\begin{align}
S&=\lambda S_{1}+\lambda^{2} S_{2}+O(\lambda^{3})
\end{align}
with 
\begin{align}
S_{1}&=-\sum_{k=0}^{p-1}(\text{Tr}X^{k})(\text{Tr}X^{p-1-k})\nonumber\\
&=-2 N(\text{Tr}X^{p-1})-\sum_{k=1}^{p-1}(\text{Tr}X^{k})(\text{Tr}X^{p-1-k})\\
S_{2}&=p\sum_{k=0}^{2p-2}(\text{Tr}X^{k})(\text{Tr}X^{2p-2-k})-\frac{1}{2}\sum_{k,l=0}^{p-1}
(\text{Tr}X^{k+l})(\text{Tr}X^{2p-2-k-l})\nonumber\\
&=p\sum_{k=0}^{2p-2}(\text{Tr}X^{k})(\text{Tr}X^{2p-2-k})-\sum_{k=0}^{p-2}(k+1)(\text{Tr}X^{k})(\text{Tr}X^{2p-2-k})-\frac{p}{2}(\text{Tr}X^{p-1})^{2}\nonumber\\
&= N(2p-1)(\text{Tr}X^{2p-2})+\sum_{k=1}^{p-2}(2p-1-k)(\text{Tr}X^{k})(\text{Tr}X^{2p-2-k})\nonumber\\
&+\frac{p}{2}(\text{Tr}X^{p-1})^{2}\label{explicitS:eq}
\end{align}

For example, with $p=3$,
\begin{align}
S_{1}&=-2 N\text{Tr}\,X^{2}-(\text{Tr}\,X)^{2}\\
S_{2}&=
5 N \text{Tr}\,X^{4}+4(\text{Tr}\,X^{3})(\text{Tr}\,X)
+\frac{3}{2}(\text{Tr} X^{2})^{2}.
\end{align}

Let us denote by $\langle\cdots\rangle$ the Gaussian averages. The partition function computed with the effective action up to order 
$O(\lambda^{3})$ reads
\begin{align}
\big\langle \exp S\big\rangle=
1+\lambda \big\langle S_{1}\big\rangle+\lambda^{2} \big\langle S_{2}\big\rangle+\frac{\lambda^{2}}{2} \big\langle (S_{1})^{2}\big\rangle+O(\lambda^{3})
\end{align}
while in conventional perturbation theory it reads
\begin{align}
\big\langle \exp -N\lambda\text{Tr}X^{p}\big\rangle=
1-\lambda N\big \langle\text{Tr}X^{p}\big\rangle
+\frac{\lambda^{2}N^{2}}{2} \big\langle (\text{Tr}X^{p})^{2}\big\rangle+O(\lambda^{3})
\end{align}
In order to check the equivalence of the two formalisms, it is convenient to use the Schwinger-Dyson equations to decrease the number of traces inside the average,
\begin{eqnarray}
\Big\langle\sum_{k+l=p_{1}-1}(\text{Tr}X^{k})(\text{Tr}X^{l})
\prod_{i=2}^{n}(\text{Tr}X^{p_{i}})\Big\rangle=-\sum_{j=2}^{n}p_{j}\Big\langle (\text{Tr}X^{p_{1}+p_{j}-1})\nonumber\\
\prod_{i=2\atop i\neq j}^{n}(\text{Tr}X^{p_{i}})\Big\rangle+N\Big\langle(\text{Tr}X^{p_{1}}).
\prod_{i=2}^{n}(\text{Tr}X^{p_{i}})\Big\rangle
\end{eqnarray}
Establishing the equality of order $\lambda$ terms is an easy task using the Schwinger-Dyson equations with $p_{1}=p$ and $p_{2}=\dots=0$,
\begin{align}
 \big\langle S_{1}\big\rangle&=
  -\sum_{k=0}^{p-1}\big\langle(\text{Tr}X^{k})(\text{Tr}X^{p-1-k}) \big\rangle=N\Big\langle(\text{Tr}X^{p})\Big\rangle.
\label{SD}
\end{align}
At order $\lambda^{2}$, we start with the term with four traces
\begin{align}
\frac{1}{2} \big\langle (S_{1})^{2}\big\rangle&=
\bigg\langle 
\frac{1}{2} \sum_{k=0}^{p-1}(\text{Tr}X^{k})(\text{Tr}X^{p-1-k})\sum_{l=0}^{p-1}(\text{Tr}X^{l})(\text{Tr}X^{p-1-l})
\bigg\rangle\nonumber\\
&=
- \sum_{l=0}^{p-1} l\bigg\langle 
(\text{Tr}X^{p-1+l})(\text{Tr}X^{p-1-l})
\bigg\rangle\nonumber\\
&+\frac{N}{2}\Big\langle(\text{Tr}X^{p}) \sum_{l=0}^{p-1}(\text{Tr}X^{l})(\text{Tr}X^{p-1-l})\Big\rangle
\end{align}
Let us use the Schwinger-Dyson equations \eqref{SD} again to reduce the three traces term 
\begin{align}
\frac{N}{2}\Big\langle(\text{Tr}X^{p}) \sum_{l=0}^{p-1}(\text{Tr}X^{l})(\text{Tr}X^{p-1-l})\Big\rangle
&=-
\frac{pN}{2}\Big\langle(\text{Tr}X^{2p-1})\Big\rangle+\frac{N^{2}}{2}\Big\langle(\text{Tr}X^{p})(\text{Tr}X^{p})\Big\rangle
\end{align}
Then, we obtain
\begin{align}
\frac{1}{2} \big\langle (S_{1})^{2}\big\rangle
&=
- \sum_{k=0}^{p-2} (p-1-k)\bigg\langle 
(\text{Tr}X^{2p-2-k})(\text{Tr}X^{k})
\bigg\rangle \nonumber\\
&-\frac{pN}{2}\Big\langle(\text{Tr}X^{2p-1})\Big\rangle+\frac{N^{2}}{2}\Big\langle(\text{Tr}X^{p})(\text{Tr}X^{p})\Big\rangle
\end{align}
Combining all contributions to order $\lambda^{2}$,we are left with
\begin{eqnarray}
\big\langle S_{2}\big\rangle+
\frac{1}{2} \big\langle (S_{1})^{2}\big\rangle&=
p\bigg\langle\sum_{k=p-1}^{2p-2}(\text{Tr}X^{k})(\text{Tr}X^{2p-2-k})\bigg\rangle-\bigg\langle\frac{p}{2}(\text{Tr}X^{p-1})^{2}\bigg\rangle\nonumber\\
&-\frac{pN}{2}\Big\langle(\text{Tr}X^{2p-1})\Big\rangle+\frac{N^{2}}{2}\Big\langle(\text{Tr}X^{p})(\text{Tr}X^{p})\Big\rangle
\end{eqnarray}
We may combine the first two terms on the RHS and use the Schwinger-Dyson equation again,
\begin{align}
p\bigg\langle\sum_{k=p-1}^{2p-2}&(\text{Tr}X^{k})(\text{Tr}X^{2p-2-k})\bigg\rangle-\bigg\langle\frac{p}{2}(\text{Tr}X^{p-1})^{2}\bigg\rangle\nonumber\\
&=
\frac{p}{2}\bigg\langle\sum_{k=0}^{2p-2}(\text{Tr}X^{k})(\text{Tr}X^{2p-2-k})\bigg\rangle
=
\frac{pN}{2}\Big\langle(\text{Tr}X^{2p-1})\Big\rangle.
\end{align}
Finally, only the term identical to the conventional perturbative one remains,
\begin{align}
\big\langle S_{2}\big\rangle+
\frac{1}{2} \big\langle (S_{1})^{2}\big\rangle=\frac{N^{2}}{2}\Big\langle(\text{Tr}X^{p})(\text{Tr}X^{p})\Big\rangle\,.
\end{align}

\subsection{Loop Vertex Expansion and the Standard Perturbation Theory for the quartic interaction case}
\label{LVESPT}

Let us give a combinatorial proof of the equivalence between our loop vertex expansion formulation and conventional perturbation theory for the quartic case ($p=2$), including terms of order $\lambda^{2}$. 
In this subsection we return to the case of rectangular matrices $N_l$ by $N_r$ for the more transparent appearance of the combinatorial factors.
Then, 
(with our convention regarding the symmetry factors and the scaling of the action in $N_{r}$ only), 
the free energy normalized by the Gaussian one reads
\begin{align}
\log Z_{\text{perturbation}}&= \log \int dMdM^{\dagger}\exp-\big\{N_{r}\text{Tr}(MM^{\dagger})+N_{r}\lambda\text{Tr}(MM^{\dagger})^{2}\big\}\nonumber\\
&=-\lambda\big\langle \text{Tr}(MM^{\dagger})^{2}\big\rangle+
\frac{\lambda^{2}}{2}\big\langle \text{Tr}(MM^{\dagger})^{2}\text{Tr}(MM^{\dagger})^{2}\big\rangle\nonumber\\
&-
\frac{\lambda^{2}}{2}\big\langle \text{Tr}(MM^{\dagger})^{2}\rangle\langle\text{Tr}(MM^{\dagger})^{2}\big\rangle+O(\lambda^{3})
\end{align}

This may be  expanded over connected oriented ribbon graphs with one or two vertices. These vertices are four-valent, with alternating incoming and out going edges. The first two terms correspond to non necessarily connected maps while the last one subtracts the disconnected part.

 Let us describe these graphs and their contributions. The contribution of faces is singled out in the last factor.

With one vertex, we have two double tadpoles (two self-loops on a single vertex), whose contribution is
\begin{align}
-\frac{\lambda}{N_{r}}\times (N_{l}+N_{r})=-\lambda\frac{N_{l}+N_{r}}{N_{r}}
\end{align}

With two vertices, we have ten maps. First the planar and non planar sunshines.
\begin{itemize}
\item Sunshine (two vertices joined by four edges, in a planar manner)
\begin{align}
\frac{\lambda^{2}}{(N_{r})^{2}}\times(N_{l}N_{r})^{2}=\lambda^{2}(N_{l})^{2}
\end{align}
\item Twisted sunshine (two vertices joined by four edges, in a non planar manner)
\begin{align}
\frac{\lambda^{2}}{(N_{r})^{2}}\times(N_{l}N_{r})=\lambda^{2}\frac{N_{l}}{N_{r}}
\end{align}
\end{itemize}

Then, there are 8 graphs obtained by joining the vertices by two lines and inserting and extra self-loop at each vertex. The latter may be inserted in several manner, on the internal or on the external faces.
\begin{itemize}
\item Insertion on the external faces
\begin{align}
\frac{\lambda^{2}}{(N_{r})^{2}}\times N_{l}N_{r}(N_{l}^{2}+N_{r}^{2})=
\lambda^{2}\frac{N_{l}(N_{l}^{2}+N_{r}^{2})}{N_{r}}
\end{align}
\item Insertion on the internal faces
\begin{align}
\frac{\lambda^{2}}{(N_{r})^{2}}\times N_{l}N_{r}(N_{l}^{2}+N_{r}^{2})=
\lambda^{2}\frac{N_{l}(N_{l}^{2}+N_{r}^{2})}{N_{r}}
\end{align}
\item One on the external and one on the internal faces
\begin{align}
\frac{4\lambda^{2}}{(N_{r})^{2}}\times (N_{l}N_{r})^{2}=
4\lambda^{2}(N_{l})^{2}
\end{align}
Thus, the logarithm of the partition function reads, in standard perturbation theory,
\begin{align}
\log Z_{\text{perturbation}}&= -\lambda\frac{N_{l}+N_{r}}{N_{r}}\nonumber\\&+\lambda^{2}\bigg(
5(N_{l})^{2}+\frac{N_{l}}{N_{r}}+2\frac{N_{l}(N_{l}^{2}+N_{r}^{2})}{N_{r}}
\bigg)
+O(\lambda^{3})
\label{pertcomb:eq}
\end{align}

To check the combinatorial coefficients, let us note that for $N_{l}=N_{r}=1$, one has (including the disconnected piece)
\begin{align}
Z=\int dzdz^{\ast}\exp-\big\{|z|^{2}+\lambda|z|^{p}\big\}=\sum_{n\geq 0}(-\lambda)^{n}\frac{(pn)!}{n!}
\end{align}
This is indeed the case as all contribution to $Z$ sum up to $2$ at order $\lambda$ and to $12$ at order $\lambda^{2}$.
\end{itemize}

In the LVE, we work with oriented trees on labelled vertices. Performing the contour integral yields decoration of the vertices with effective actions;
 Each vertex labelled $i$ carries an effective action $S(X_{i})$ ($X_{i}=M_{i}M_{\dagger}$), which writes, in the quartic case (see \eqref{explicitS:eq} for $p=2$)
\begin{align}
S(X)&=-\lambda(N_{l}+N_{r})(\text{Tr}X) +\lambda^{2} 
\Big\{\frac{3}{2}(N_{l}+N_{r})(\text{Tr}X^{2})+(\text{Tr}X)^{2}\Big\}
+O(\lambda^{3})
\end{align}
At order $\lambda$, there is only the empty tree with a single vertex labelled 1. Its contribution reads
\begin{align}
\lambda(N_{l}+N_{r})\big\langle\text{Tr}(M_{1}M^{\dagger}_{1})\big\rangle_{1}&=-\lambda N_{l}(N_{l}+N_{r})^{2}
\end{align}
with the normalised Gaussian average
\begin{align}
\big\langle\cdots\big\rangle_{1}&=\int dM_{1}dM^{\dagger}_{1} (\cdots) \exp-N_{r}\text{Tr}(M_{1}M^{\dagger}_{1})
\end{align}
At order $\lambda^{2}$, the LVE includes oriented trees with one and two vertices. 

\begin{itemize}
\item Tree with a single vertex decorated with the order $\lambda^{2}$ effective action and a Gaussian average over a single matrix
\begin{align}
\lambda^{2}\Big\langle\frac{3}{2}(N_{l}+N_{r})(\text{Tr}X^{2})+(\text{Tr}X)^{2}\Big\rangle_{1}&=
\frac{3}{2}\lambda^{2}\frac{N_{l}(N_{l}+N_{r})^{2}}{N_{r}}
+\lambda^{2}N_{l}^{2}+\lambda^{2}\frac{N_{l}}{N_{r}}
\end{align}
\item Trees with two vertices 1 and 2 and an oriented edge, either from 1 to 2 or from 2 to 1, the two vertices being decorated with the effective action at order 1,
\begin{align}
\frac{\lambda^{2}}{2}(N_{l}+N_{r})^{2}\int_{0}^{1}dx\big\langle\text{Tr}(M_{1}M_{2}^{\dagger})+\text{Tr}(M_{2}M_{1}^{\dagger})\big\rangle_{12}&=\frac{\lambda^{2}}{2}\frac{N_{l}(N_{l}+N_{r})^{2}}{N_{r}}
\end{align}
with the Gaussian measure on two matrices  of covariance $C_{12}$
\begin{align}
\big\langle\cdots\big\rangle_{12}&=\int d\mu_{C_{12}}(M_{1},M^{\dagger}_{1},M_{2},M^{\dagger}_{2})\, (\cdots),\qquad
C_{12}=\frac{1}{N_{r}}\begin{pmatrix}1&x\\x&1\end{pmatrix};
\end{align}
\end{itemize}
Therefore, the LVE expansion  yields
\begin{align}
\log Z_{\text{LVE}}&= -\lambda\frac{N_{l}+N_{r}}{N_{r}}+\lambda^{2}\bigg(
2\frac{N_{l}(N_{l}+N_{r})^{2}}{N_{r}}
N_{l}^{2}+\frac{N_{l}}{N_{r}}
\bigg)
+O(\lambda^{3}).
\label{LVEcomb:eq}
\end{align}
Comparing with the perturbative expansion \eqref{LVEcomb:eq}, we see that $\log Z_{\text{LVE}}=\log Z_{\text{perturbative}}$.



\end{document}